\documentclass[nohyper]{JHEP3}
\usepackage{amsmath}
\usepackage{amssymb}
\usepackage{cite}
\usepackage{xspace}

\usepackage{epsfig}

\psfull

\newcommand{\ie}{\emph{i.e.}\ }
\newcommand{\eg}{\emph{e.g.}\ }
\newcommand{\cnf}{\emph{cf.}\ }

\def\cO#1{{\cal{O}}\left(#1\right)}

\def\vn{\vec{n}}

\def\hc{{\cal H}_{\mathrm{C}}}
\def\hr{{\cal H}_{\mathrm{R}}}

\def\ec{{\cal E}_{\mathrm{C}}}
\def\elim{{\cal E}_{\mathrm{lim}}}

\def\qbar{{\bar q}}

\def\al{\alpha}
\def\be{\beta}

\def\cM{{\cal{M}}}    
\def\cP{{\cal{P}}}    
\def\cS{{\cal{S}}}    
\newcommand{\cV}{{\cal V}}

\def\cR{{\cal{R}}}               

\def\MSbar{\overline{\mbox{\scriptsize MS}}}

\def\NC{N_{\textsc{c}}}
\def\cf{C_F}
\def\CF{\cf}

\def\ca{C_A}
\def\nf{n_{\!f}}

\def\as{\alpha_{{\textsc{s}}}}

\def\asb{{\bar \alpha}_{{\textsc{s}}}}

\newcommand{\order}[1]{\mathcal{O}\left(#1\right)}

\newcommand{\prll}{{\parallel}}
\newcommand{\dispatch}{\texttt{dispatch}}
\newcommand{\disent}{\textsc{disent\xspace}}
\newcommand{\disaster}{\textsc{disaster}\texttt{++}\xspace}

\newcommand{\bC}{\boldsymbol{C}}

\newcommand{\bcD}{\boldsymbol{\mathcal{D}}}
\newcommand{\cD}{\mathcal{D}}
\newcommand{\bq}{\boldsymbol{q}}

\def\ee{e^+e^-}

\title{Resummed event-shape variables in DIS.}%

\author{Mrinal Dasgupta \\
  DESY, Theory Group, Notkestrasse 85, Hamburg, Germany.}
\author{Gavin P. Salam \\
LPTHE, Universit\'es P. \& M. Curie (Paris VI) et Denis Diderot
  (Paris VII), Paris, France.}

\abstract{We complete our study of resummed event-shape distributions
  in DIS by presenting results for the class of observables that
  includes the current jet mass, the $C$-parameter and the thrust with
  respect to the current-hemisphere thrust axis.  We then compare our
  results to data for all observables for which data exist, fitting for $\as$
  and testing the universality of non-perturbative $1/Q$ effects. A
  number of technical issues arise, including the extension of the
  concept of non-globalness to the case of discontinuous globalness;
  singularities and non-convergence of distributions other than in the
  Born limit; methods to speed up fixed-order Monte Carlo programs by
  up to an order of magnitude, relevant when dealing with many $x$ and
  $Q$ points; and the estimation of uncertainties on the predictions.}


\preprint{DESY--02--104\\
  LPTHE--02--040\\
  hep-ph/0208073 \\
  August 2002\\
}

\begin{document}

\section{Introduction}

Quantum chromodynamics is special among the theories of the standard
model in that its strongly and moderately coupled domains are within
reach of experiment, and indeed unavoidable at current energies. This
leads to a wealth of non-trivial phenomena involving multiple particle
production, divergent perturbative series and large non-perturbative
effects even at formally perturbative scales.  A full understanding of
such phenomena remains one of the main aims of QCD, both because of
their fundamental interest and because they are inextricably present
in any measurement involving hadrons, be it a determination of the
properties of QCD itself or even searches for new physics.

A feature of QCD is that the more exclusive an observable, the more
difficult it is to predict. This is because more exclusive observables
often involve several disparate scales, and higher-order terms in the
perturbative series have coefficients which are enhanced by (large)
logarithms of ratios of scales. Furthermore whenever the observable
involves integrals over phase space which extend significantly into
the infrared, or when one of the scales is close to the
strong-coupling region, the observable is likely to be subject to
large non-perturbative contributions.

A class of observables involving an interesting compromise between
infrared sensitivity and perturbative calculability is that of event
shapes and their distributions. These infrared-collinear (IRC) safe
observables measure properties of the flow of energy-momentum, for
example the normalised squared invariant mass of a hemisphere, $\rho$.
When pushing such observables to the exclusive limit (in the case of
$\rho$, small invariant masses) the perturbative distribution develops
double logarithms at each order of perturbation theory $(\as \ln^2
1/\rho)^n$, which must be resummed to all orders. Furthermore because
such observables are usually linear in soft momenta, there are
significant non-perturbative effects, formally of order $1/Q$, which
cannot be ignored.

This class of observables has already been widely examined for $2$-jet
events in $\ee$ collisions and resummations of large-logarithmic terms
\cite{CTTW,CTW,eeJets,CW,BroadResum,BanfiSalamZander} have been in
existence for some time now. The literature is at first sight slightly
confusing when it comes to defining the accuracy of these
resummations. Let us use $R(\rho)$ to denote the probability that the
jet mass is smaller than some value $\rho$. Then we can write $R$ as
\begin{equation}
  R(\rho) = 1 + \sum_{n=1} \as^n \left(\sum_{m=0}^{2n} R_{nm} \ln^m
    \frac1\rho + \order{\rho}\right)\,.
\end{equation}
One can define leading logarithms (LL) as consisting of all terms 
with $m = 2n$,
next-to-leading logarithms (NLL) as terms with $m = 2n-1$, and so
on (\eg \cite{eeJets,DISJets}). In this terminology, the state of the art
for resummations is NNLL
order, and in terms of relative accuracies on $R$ it means that for
$\as L^2 \sim 1$, $R$ is determined to within relative corrections of
order $\as^{3/2}$.

For $\as L^2 \gg 1$ however, any fixed order $p$ of logarithmic
resummation in this hierarchy, N$^{p}$LL, loses its predictive power,
because there will always be neglected terms $\as^n L^{m}$, with $2n-p
> m > n$,
which are formally larger than $1$. Fortunately many observables can
be shown to exponentiate, which means that we are entitled to write
\begin{equation}
  R(\rho) = \exp\left(\sum_{n=1} \as^n \left(\sum_{m=0}^{n+1} G_{nm} \ln^m
    \frac1\rho + \order{\rho}\right)\right)\,,
\end{equation}
(in some cases $R$ needs to be broken up into a sum of terms, each of
which exponentiates, but with different exponents). With this form of
expansion, leading logarithms now refer to all terms $m=n+1$, while
next-to-leading logs (or single logs --- SL) are those with $m=n$
(\eg \cite{CTTW}). The
state of the art in this form of expansion is NLL accuracy. It means
that for $\as L \sim 1$ the neglected terms correspond to
modifications of $R$ of relative order $\as$. More generally, for
observables that do
not exponentiate, one can use this condition on the relative size of
higher-order corrections to define NLL accuracy.

There is partial overlap between the terms needed for NNLL accuracy in
$R$ and NLL accuracy in $\ln R$. Our aim is to guarantee both, though
unless we explicitly specify otherwise, our statements of 
accuracy will refer to the expansion of $\ln R$.

With the 2-jet NLL $\ee$ calculations, there have been extensive
comparisons to data (see for example \cite{LEPdist}) and much has been
learnt from these studies, especially regarding our understanding of
hadronisation effects, for which several related theoretical
approaches (\eg \cite{ManoharWise,Webber94,DW,DMW,AK,BB,KS,DGE}),
mostly based on renormalons \cite{renormalon}, have been developed and
tested.

Of course, event-shape observables are not restricted to $\ee$
collisions and
there are good reasons to extend their study to other processes. Many
of the hadronisation models embody a concept of \emph{universality},
whereby all $1/Q$ hadronisation effects should be governed by a
restricted set of non-perturbative parameters which are both
observable and process independent. In $\ee$ event-shapes there is
good evidence in favour of universality \cite{UniversalityTest}, but in
order to be fully established it needs to be tested in processes
involving incoming hadrons as well. It is also important that the
perturbative predictions for exclusive final-state properties be
tested in a wider context than just $\ee$, both so as to establish
confidence in their 
general applicability and in view of the fact that they may well come
to play a role in searches for new physics through studies of the
radiation pattern associated with the different colour flows that are
relevant in `new physics' processes as opposed to the QCD
backgrounds \cite{Associated}.

It was to investigate these issues in more detail that a study of NLL
resummed
DIS Breit-frame current-hemisphere ($\hc$) event-shape observables was
initiated in \cite{ADS,ADSdurham,dassalbroad}.  From the point of view
of studying hadronisation effects, DIS observables are particularly
interesting in that there is a considerable amount of data
\cite{H1OldData,H1NewData,ZEUS} over a wide range of $Q$ values
(within the same 
experiment), making it possible to explicitly test the predicted
$Q$-dependence of hadronisation effects.  This was part of the
motivation for the original theoretical studies \cite{DasWeb,DasWebMi}
of hadronisation effects in mean values of DIS event shapes.

The DIS resummations carried out in \cite{ADS,dassalbroad} dealt
with observables 
defined with respect to the boson ($z$) axis --- two variants of the
thrust ($\tau_{zQ}$ and $\tau_{zE}$) and one of the jet broadening
($B_{zE}$). In the current paper we complete our study of resummed DIS
event-shapes by presenting results for three observables defined
independently of the boson axis (except through the restriction that
only particles in $\hc$ are measured) and by comparing all of our
resummed results to data.\footnote{Except for $\tau_{zQ}$ for which no
  data currently exist.}

The three observables resummed in this paper are the thrust with
respect to the thrust axis and normalised to the current hemisphere
energy, $T_{tE}$, the invariant squared jet mass, $\rho_E$ and the
$C$-parameter, $C_E$. Their exact definitions are as follows:
\begin{equation}
\label{eq:thrust}  
T_{tE} = \max_{\vn} \frac {\sum_{\hc}  |\vec{P_i}.\vn|}{\sum_{\hc}
  |\vec{P_i}|}\,, 
\end{equation}
\begin{equation}
  \label{eq:rhodef}
\rho = \frac {\left(\sum_{\hc}  P_i\right)^2}{4 \left(\sum_{\hc}
      |\vec{P}_i|\right)^2 }
\end{equation}
\begin{equation}
\label{eq:Cparam}
C_E = \frac{3}{2}  
\frac{\sum_{a,b \in \hc}|\vec{P_a}||\vec{P_b}| \sin^2
  \theta_{ab}}{\left (\sum_{\hc} 
      |\vec{P_i}|\right)^2}
\end{equation}
Additionally, we refer to $\tau_{tE} = 1-T_{tE}$. Events where the
energy in the current hemisphere is less than some value $\elim$ are
discarded. This is necessary to ensure the all-order infrared and
collinear (IRC) safety of the observable, in particular to protect
against divergences associated with events containing hard emissions
in the remnant hemisphere ($\hr$) alone and arbitrarily soft emissions
in $\hc$.  It is important that $\elim$ not be too small a fraction of
the photon virtuality $Q$, in order to avoid developing large $\ln
\elim/Q$ terms.

The fact that the above observables do not involve the boson axis in
their definitions means that they are essentially insensitive to
radiation in the remnant hemisphere --- this is in contrast to
the boson axis observables considered previously, which were all
sensitive to radiation in $\hr$ through recoil effects on the current
quark. This has several consequences for the resummation: the
double-logs are less strong because they only involve the suppression
of soft and collinear radiation from one hard parton; a fraction $\as$
of the time, the hard parton in $\hc$ is not a quark but a gluon,
implying different colour factors for the double-logs; at the single
log-level the appropriate factorisation scale for the parton
distributions is $Q^2$, independently of the value of the observable;
and the resummation has non-global single-logs, discovered recently in
\cite{dassalNG1,dassalNG2} which involve multiple energy-ordered
large-angle gluons in $\hr$ emitting coherently into $\hc$.

The different aspects of the resummation are discussed in
sections~\ref{sec:kinematics}, \ref{sec:resummation} and
\ref{sec:discont} with the latter section addressing also the issue of
previously neglected non-global logarithms for the boson-axis
observables. This leads to our introducing the concept of
\emph{discontinuously global} variables such as $\tau_{zE}$ and
dynamically discontinuously-global variables such as $B_{zE}$. In
practice the phenomenological impact of the non-global logarithms on
these observables is small.

In addition to there being a need for a resummation in the single
current-jet limit ($V \to 0$), 
it turns out that there are poorly convergent, log-enhanced
structures in other regions of phase space, associated with
discontinuities or non-smoothness in low-order distributions. These
are analogous to the problems seen for the $\ee$ $C$-parameter at
$C=3/4$, and a
resummation generally leads to a smoothing-out of these structures
\cite{CataniWebberInternalDiv}. In most cases their impact is
restricted to a relatively small region, so in
section~\ref{sec:nonsmooth} we limit ourselves to illustrating the
origin of the problems for the the DIS $C_E$ parameter, and giving
a list of the non-smooth structures that arise for the other
observables. We do not carry out any resummation of these observables around 
the singular structures, restricting ourselves to resummation near the 
exclusive boundary, $V \to 0$, where $V$ is the event shape in question.

Another, technical, issue that needs to be dealt with is that of
determining the fixed-order distributions for all the observables.
This turns out to be a very computer-intensive process if one wishes
to obtain a reasonable precision and resolution. So in
section~\ref{sec:speedyMC} we present a method for speeding up
fixed-order Monte Carlo programs such as \disent \cite{DT} or
\disaster \cite{DR} through the reuse of the same event to calculate
many $x$ and $Q^2$ points at the same time.

Once one has the resummed and fixed-order predictions (matched with
the procedures of \cite{CTTW,dassalbroad}) one can then address
(section~\ref{sec:NP}) the issue of including non-perturbative
corrections. We follow the approach of \cite{DW,DMW,DokWeb97} based on
the concept of a universal infrared-finite coupling, where the power
correction is essentially a shift of the distribution by an amount
which depends on a moment $\alpha_0$ of the coupling in the infrared
and on a calculable, observable-dependent coefficient (as determined
in \cite{DasWeb,DasWebMi,dassalbroad}).

With all the ingredients in place we then show, in
section~\ref{sec:impact} the impact of the resummation, and
discuss choices and uncertainties associated with unknown
higher-orders and with certain prescriptions for the non-perturbative
effects.

This is followed in section~\ref{sec:data} by comparisons to data, and
fits for $\as$ and $\alpha_0$. Finally, in
section~\ref{sec:conclusions} we give our conclusions, summarising
some of the many developments made in the course of our DIS
resummation project.

\section{Kinematics}
\label{sec:kinematics}

Here we establish the kinematic properties of the different variables
in which we are interested. The expressions derived will be
approximate but sufficient for single-log/next-to-leading log (NLL)
accuracy.

In order to achieve this precision we need to consider configurations
of soft and/or collinear gluons in addition to the hard initiating
parton (quark or gluon).  To be precise we want expressions for the
observable in several kinematic configurations:
\begin{itemize}
\item A hard particle in $\hc$, close to the boson axis, accompanied
  by any number of soft and collinear gluons. Here the expression for
  the observable has to be correct to within terms of relative order
  of a power of the softness so as to account for both double and
  single logs.
  
\item A hard particle (close to the boson axis) in $\hc$ along with an
  energetic almost collinear gluon. This region is only associated
  with single-logs and the expression for the observable need simply
  be correct to within an overall factor (because replacing $\as^n
  \ln^n 1/\rho$ by $\as^n \ln^n A/\rho$, with $A$ some fixed number,
  is equivalent to a NNLL change).
  
\item A hard particle (close to the boson axis) in $\hc$ accompanied
  by a soft wide-angle gluon in $\hc$. This situation is also
  responsible only for single logs and so it is again sufficient to
  have an expression for the observable correct to within an overall
  factor.
  
\item A hard particle in $\hc$ away from the boson axis,
  accompanied by any number of particles soft and collinear to the
  hard parton. This kind of hard configuration is suppressed by a
  power of $\as$ and so only gives NNLL terms (both in $R$ and in $\ln
  R$). There it is sufficient to have an expression for the observable
  valid to within an overall factor.
  
\end{itemize}
In addition, for our resummation approach to be valid, the conditions
for exponentiation (discussed in \cite{dassalbroad,BanfiSalamZander})
must also hold. This is the case for all the observables discussed
here.

What we shall actually do is derive expressions for the jet mass,
which has already been considered in various related cases in
\cite{CT,CMW,dassalNG1}, and then show that to the accuracy that we
need, other observables are related to it by a constant factor.

We first obtain an expression for $\rho$ relevant for many soft (and
optionally 
collinear) emissions. We take a hard parton $p$, close to the photon
axis and with energy $p_0 \simeq Q/2$, and any number of soft and
collinear particles $k_i$, with energies $\omega_i$ and angles
$\theta_{ip}$ to the hard parton. We denote by $E = p_0 +
\sum_{i\in\hc}\omega_i$ the total energy of all particles in $\hc$.

With this kinematics, the jet mass is given by
\begin{equation}
  \rho = \frac{1}{4 E^2} \left(p+\sum_{i\in \hc}
  k_i\right)^2 \simeq 
  \frac{1}{Q} \sum_{i\in \hc}\omega_i (1-\cos \theta_{ip}) \,,
\end{equation}
To extend this result to the case of a single hard collinear emission,
$k_1$, we need to take into account the recoil of the hard parton and
the result for $\rho$ is
\begin{equation}
  \rho  \simeq
  \frac{\omega_1}{Q} \left(1 -
    \frac{2\omega_1}{Q}\right) (1-\cos\theta_{1p}) \,.
\end{equation}
This is the information used for example in the resummations of
\cite{CT,CMW,dassalNG1}.

An expression for the thrust in terms of the jet mass, in the limit
where all emissions are soft (and at a polar angle of less than
$\pi/2$ from the quark direction) and/or collinear, can be obtained by
noting that in this limit the thrust axis coincides with the direction
given by the sum of all three-momenta in the hemisphere, as in the
$\ee$ case \cite{CTTW}.

Denoting by $\vec P$ the total vector $3$-momentum sum, we can rewrite
$\rho$ as
\begin{equation}
  \label{eq:rhoGen}
  \rho = \frac{E^2 - |\vec P|^2}{4E^2}\,,
\end{equation}
while the thrust is given by
\begin{equation}
  \label{eq:tauGen}
  \tau_{tE} = 1 - \frac{1}{E|\vec P|}\left( |{\vec p} . {\vec P}|
     + \sum_{i \in \hc} |{\vec k}_i . {\vec
       P}|\right)
   = \frac{E - |\vec P|}{E}
\end{equation}
where we have made use of the condition that none of the particles is
moving backwards with respect to $\vec P$.

In the limit $\rho \ll 1$, implying $E - |\vec P| \ll E$, we
immediately see that $\tau_{tE} = 2\rho_E + \order{\rho^2}$, allowing
us to reuse the $\rho$ resummation for $\tau_{tE}$, simply replacing
every occurrence of $\rho_E$ with $\tau_{tE}/2$.\footnote{Strictly
  speaking, for this to be true one also needs the easily demonstrated
  property that $\tau_{tE} \sim \rho_E$ when the hard particle is at
  large angles (and emitted gluons can move backwards with respect to
  the thrust axis and still be in $\hc$).}

In the case of the $C$-parameter we need to explicitly derive
expressions for it before being able to compare to the jet mass.
Starting from eq.~\eqref{eq:Cparam} we see that for any number of soft
(and optionally collinear) emissions:
\begin{equation}
C_E \simeq 6 \sum_{i\in \hc}\frac{\omega_i}{Q} \sin \theta_{1p}^2,
\end{equation}
while for a single hard collinear emission, $k_1$, we have to take
into account the recoil of the hard parton to get
\begin{equation}
C_E \simeq 6 \frac{(1 - 2\omega_1/Q)\omega_1}{Q} \sin \theta_{1p}^2.
\end{equation}
Comparing to the corresponding expressions for the jet-mass we see
that in the presence of collinear emissions (many soft or one hard)
$C_E = 12 \rho_E$. For large-angle soft emissions we do not have an
exact proportionality, but both $\rho_E$ and $C_E$ receive
contributions proportional to $\omega_i$ and so should have the same
large-angle single logs. Therefore the resummed result for $C_E$ is
simply that for $\rho_E$ with the replacement $\rho_E \to C_E/12$. 

One final point to note is that because $\rho_E$ and $\tau_{tE}$ are
proportional in all situations involving a single soft or collinear
emission in $\hc$, the corresponding pure $\order{\as}$ contributions
to the resummed results (the $C_1$ terms discussed below) are
identical, and the
coefficients of the power corrections are also related by the same
proportionality factor. In contrast, because of the different
dependence on large-angle soft emissions, for $C_{E}$ one of the
constant terms differs (by a $\delta$-function in $x$), as discussed
in appendix~\ref{app:ConstantPieces} and the power correction is not
$12$ times larger than that for $\rho$, but rather only $3\pi$ times
larger.

\section{Resummation}
\label{sec:resummation}

To write down a resummed result to NLL accuracy for the current
hemisphere jet mass $\rho$ we simply refer to Ref.~\cite{dassalNG1}.
There we established a formula for the hemisphere jet mass in
$e^{+}e^{-}$ annihilation and an essentially identical form applies
here. We define the integrated current hemisphere jet mass
distribution at a given $x,Q^2$,
\begin{equation}
  \label{eq:Rdef}
  R(\rho,x,Q^2) = 
  \left(\frac{d\sigma_0}{dx dQ^2}\right)^{-1} \int_0^\rho
  \frac{d\sigma}{d\rho'dx dQ^2} d\rho'\,,
\end{equation}
where $d\sigma_0/dxdQ^2$ is the Born cross section.\footnote{We could
  equally well have normalised to the total cross section, but this
  would lead to the $n^\mathrm{th}$ order perturbative expansion for
  $R$ involving combinations of parton distributions such as $(C_a
  \otimes q) \ldots (C_n \otimes q) / [q(x,Q^2)]^n$. It is thus
  simpler to calculate $R$ defined as normalised to the Born cross
  section and then to renormalise it as appropriate when comparing to
  data.} %
This integrated jet-mass distribution is given to NLL accuracy by the
expression
\begin{equation}
\label{eq:result}
R(\rho,x,Q^2) = \left (1+\asb C_1^q(x,Q^2)\right) S(\as L)
\Sigma_q(\as, L)
+\asb C_1^g(x,Q^2) \Sigma_g(\as, L)\,, \quad L \equiv \ln
\frac1\rho\,.
\end{equation}
The precise analytical forms for the different elements in this
equation are given in the appendices. Here we limit ourselves to an
explanation of the physical meaning of the various factors.

The first term in eq.~\eqref{eq:result} accounts for situations in
which the hard particle in $\hc$ is a quark; $\Sigma_q(\as, L)$ is the
form factor for the hemisphere mass to be less than $e^{-L}$ assuming
just independent eikonal emission off a $q{\bar q}$ line (including
hard collinear corrections). It can be written as
\begin{equation}
\label{eq:CataniJetq}
\Sigma_{q}(\as,L) = \int_0^{e^{-L}} J_{q}\!
\left(\as,\frac{k^2}{Q^2}\right) \frac{dk^2}{k^2}\,,
\end{equation}
in terms of the jet-mass distribution, $J_{q}$, computed several years
ago \cite{CT,CMW}. It contains both double and single logarithms.

The factor $\cS(\as L)$ embodies the single-logarithmic corrections
associated with the fact that $\rho_E$ is non-global, namely that it
measures emissions in $\hc$ only. This means that a constraint on
$\rho_E$ does not imply any explicit constraint on emissions in $\hr$, and
large-angle gluons in $\hc$ are actually radiated off an ensemble of
energy-ordered large-angle gluons in $\hr$. Thus the pattern of
large-angle gluon radiation in $\hc$ differs from independent emission
off an eikonal $q\bar q$ line at the single logarithmic level, and
this needs to be accounted for with the factor $\cS(\as L)$, as
discussed in more detail in section~\ref{sec:discont} (together with
related, but rather subtle contributions that were neglected in our
previous work \cite{ADS,dassalbroad}). It is to be noted that beyond
order $\as^2 L^2$, $\cS(\as L)$ is known only numerically and in the
large-$\NC$ limit.

The $\rho$-independent $\asb C_1^q(x,Q^2)$ contribution is the only
piece in the first
term of \eqref{eq:result} that is sensitive to the parton
distribution functions (PDFs). It has a number of separate, though not
always separately well-defined, origins. One physically identifiable
contribution to $C_1^q$ is from the configuration where one has a lone
hard quark in $\hc$ and a gluon in $\hr$. This on its own is divergent
and cancels against corresponding virtual corrections and against the
term from the factorisation of the parton distribution functions
(PDFs).  This implies that $C_1^q$ has both factorisation scheme and
scale dependence, similar to what is discussed in appendix A of
\cite{ADS} (Eqs.~(A.9) and (A.10)). It should be kept in mind that
when there is a hard large-angle gluon in $\hr$, then the hard quark
in $\hc$ will not be aligned along the boson axis nor will it have
energy $Q/2$, and a number of the approximations made at single log
level in $\Sigma_q$ and $\cS$ will be incorrect --- however this is
beyond our accuracy, because it is relevant only a fraction $\as$ of
the time.

The remainder of $\as C_1^q$ is associated with approximations made in
$\Sigma_q$ for the pattern of real and virtual corrections for a hard
gluons and (in some cases) with approximations for the dependence of
the observable on soft large-angle and hard collinear emissions. We
note that configurations with an empty $\hc$ (strictly ${\cal E_C}
< \elim$) are considered as having an undefined value of the jet mass,
and so are not included in the integral for $R$ and correspondingly in
the definition of $C_1^{q}$.

A final point to note about the $\as C_1^q$ piece is that when
considering $\as L \sim 1$, as we usually do, and defining our
hierarchy in terms of the series expansion of $\ln R$, the $C_1$ terms
are not formally needed to NLL accuracy. In contrast if we are in the
region $\as L^2 \sim 1$ and looking at the expansion of $R$ to NNLL
accuracy, then the $C_1$ terms are needed.

Let us now come to the second term of eq.~\eqref{eq:result}. This
accounts for the situation in which the hard parton in $\hc$ is a
gluon.  This happens in only a fraction $\as C_1^g(x,Q^2)$ of events,
and the form factor $\Sigma_g(\as, L)$ (also given in \cite{CT,CMW})
is related at leading-log level to $\Sigma_q(\as,L)$ by replacing
$\CF$ with $\ca$ in the exponent. At single log level in $\Sigma_g$
the situation would be more complicated because we cannot make the
approximation of the gluon being aligned along the boson axis or of
its having energy $Q/2$. However as in the case of $C_1^q$, given that
this whole term is multiplied by $\as$, to our overall level of
accuracy we are free to mistreat single-logs in $\Sigma_g$. In
practice we systematically neglect all single-log contributions in
this term, including those from non-global effects (which could have
been represented by an extra factor $\cS_g$ in \eqref{eq:result}).

Finally we recall that in order to derive a result for the $C_E$ and
$\tau_{tE}$ distributions one just replaces $\rho$ by $C/12$ and
$\tau_{tE}/2$ respectively throughout eq.~\eqref{eq:result} --- the
$\Sigma_q$, $\Sigma_g$ and $\cS$ functions have identical forms for
all the above variables.  It should however be noted that while the
coefficient function $C_1^q$ is the same for the $\tau_{tE}$ and
$\rho$ variables, it is different for the $C$ parameter (see
appendix~\ref{app:ConstantPieces}).

\section{Non-global logarithms and discontinuously global observables}
\label{sec:discont}

In a previous paper we introduced the concept of a non-global
observable, defined as an observable sensitive to emissions only in a
restricted angular portion of phase space. We pointed out that at
single-logarithmic accuracy, such observables receive a set of
contributions associated with energy-ordered large-angle emissions,
termed non-global logarithms \cite{dassalNG1}. In this section, we
wish to refine that definition.

\subsection{Non-global logs}
\begin{figure}[t]
  \begin{center}
    \epsfig{file=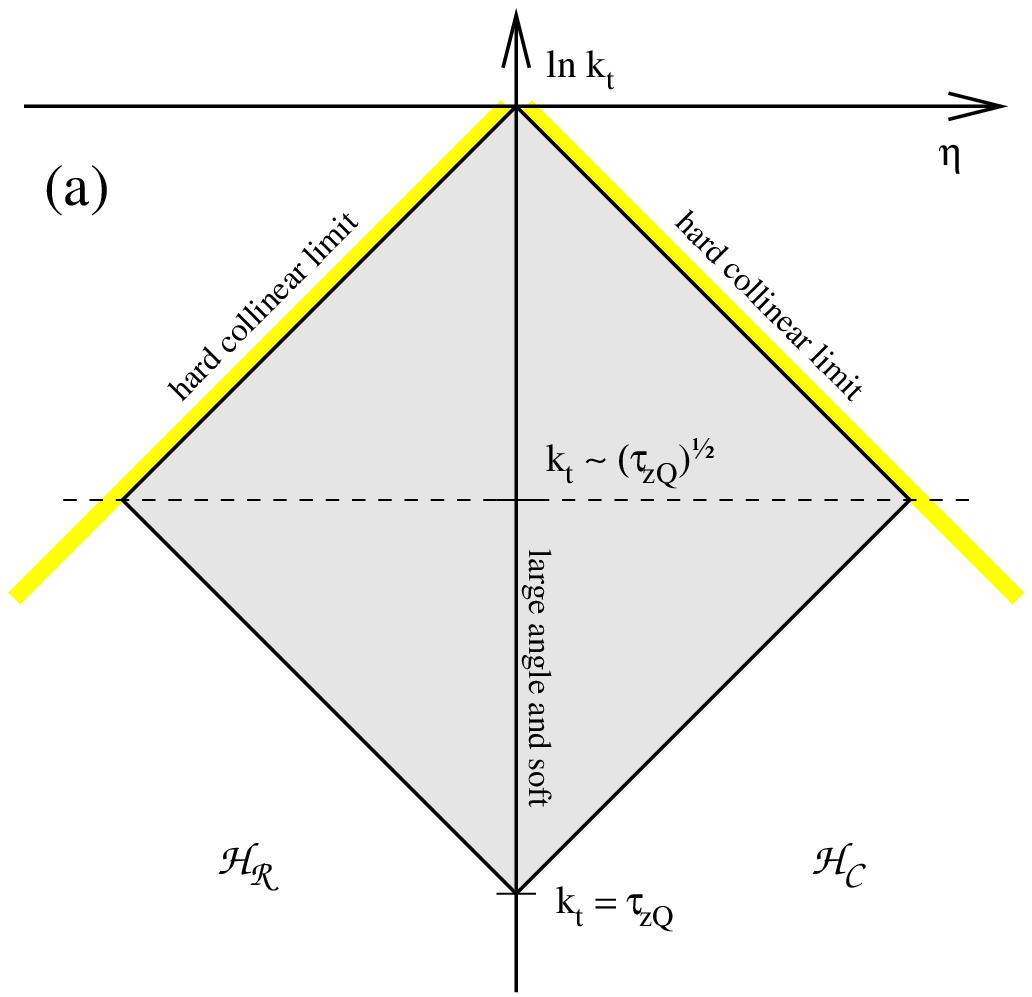,width=0.455\textwidth}\hfill
    \epsfig{file=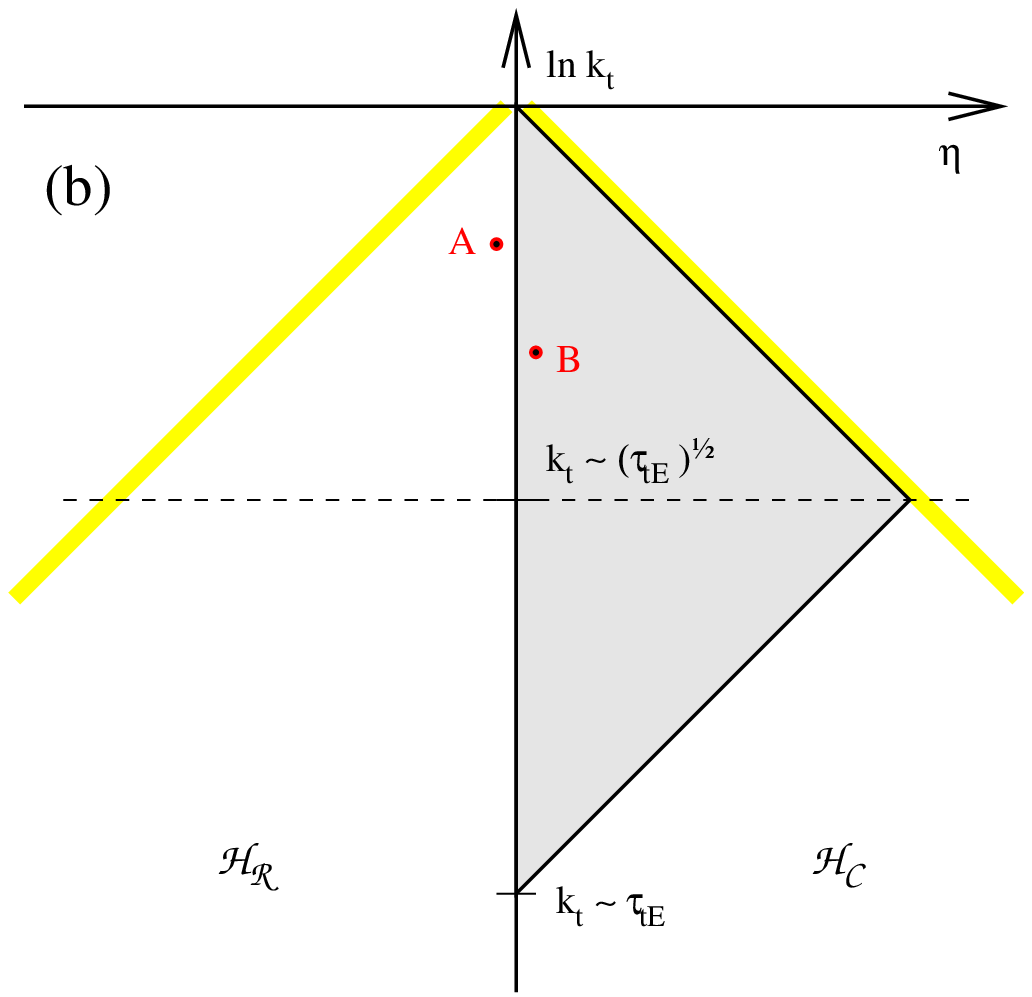,width=0.455\textwidth}
    \caption{Schematic depiction of the region in rapidity and log
      transverse momentum, in which emissions must be forbidden (grey
      shaded area) in order for an observable not be be larger than a
      given value $\tau$.  Left: shown for $\tau_{zQ}$, the
      photon-axis thrust normalised to $Q$ (a global observable).
      Right: shown for $\tau_{tE}$, the thrust-axis thrust normalised
      to the energy in $\hc$ (a non-global observable).  The yellow
      (light grey) diagonal bands indicate the hard-collinear limits.}
    \label{fig:discont}
  \end{center}
\end{figure}

The difference between a global observable and a non-global one is
illustrated in fig.~\ref{fig:discont}, which shows in grey the
phase-space regions where emissions are forbidden if the observable's
value is to be below a given value $\tau$. In the left hand figure we
consider the global observable $\tau_{zQ}$, whose value for a given
ensemble of emissions is
\begin{equation}
\label{eq:thrustglo}
\tau_{zQ} \simeq \sum_i k_{t,i}e^{-|\eta_i|}\,,
\end{equation}
where the sum $i$ runs over all soft (and collinear) particles and the
transverse momenta are implicitly normalised to $Q$ for brevity. For
observables
like $\tau_{zQ}$ it is safe to neglect energy-ordered large-angle
contributions as can be seen by considering the case with two
emissions $A$ and $B$, with $k_{t,A} \gg k_{t,B}$ and $\eta_{A} \simeq
\eta_B$: the modification of the observable's value by emission $B$ is
negligible, $k_{t,A} \simeq k_{t,A} + k_{t,B}$, and there will be full
cancellation (at the single-log level) between real and virtual
contributions for $B$.

A non-global observable is one such as $\tau_{tE}$, illustrated in
fig.~\ref{fig:discont}b. The observable's value is given by (\cnf
eq.~\eqref{eq:tauGen}) 
\begin{equation}
\label{eq:thrustnglo}
\tau_{tE} \simeq \sum_i 2k_{t,i}e^{-|\eta_i|} \Theta(\eta_i)\,,
\end{equation}
\ie there is only sensitivity to emissions in the current hemisphere,
and placing a constraint on the value of the observable implies no
direct constraint on emissions in $\hr$.\footnote{In writing the above
  formula for $\tau_{tE}$ we have neglected configurations in which
  there is a hard parton (quark or gluon) at a large angle to the the
  boson axis.  This is because the overall contribution from
  non-global logs in such situations is $\order{\as^{n+1} L^{n}}$
  (including the configuration probability $\order{\as}$) and hence
  subleading.}

In fig.~\ref{fig:discont}b,
this translates to only the current-hemisphere region being shaded.
Now let us consider the situation with emissions $A$ and $B$ as shown
in the figure, $k_{t,A} \gg k_{t,B}$ and both emissions at large
angles, with $A$ in $\hr$ and $B$ in $\hc$. Emission $A$ does not
contribute to the observable. Emission $B$ would contribute, since it
is in the shaded region --- so it must be forbidden, corresponding to
a resummation of virtual corrections. However the form factor
$\Sigma_q$ in eqs.~\eqref{eq:result} and \eqref{eq:CataniJetq} is derived
with the assumption that the pattern of soft large-angle emission is
that off an eikonal $q{\bar q}$ current, whereas the presence of $A$
at large angles in $\hr$ modifies the pattern of large-angle soft
emissions in $\hc$ (and the corresponding virtual corrections) to be
that from a $q{\bar q} g$ antenna.

So it is necessary to correct $\Sigma_q$ for the fact that the
emission pattern is not that from a $q{\bar q}$ pair, but in general
that from an ensemble that can contain any number of energy-ordered
large-angle gluons in $\hr$. At the two gluon level shown in
fig.~\ref{fig:discont}b the correction to $\Sigma_q$ is of the form
\begin{equation}
  \label{eq:twogluonNG}
   -\cf\ca \frac{\pi^2}{3} \left( \asb \int_\tau^1
    \frac{dk_t}{k_t} \right)^2\,,
\end{equation}
where the double integration over transverse momentum is associated
with the relevant $k_t$ ranges of $A$ and $B$.

At all orders, the contribution from non-global effects can be written
as a function $\cS(t)$ where $t$ is given by
\begin{equation}
  \label{eq:tdefNG}
  t = \int_\tau^1 \frac{d k_t}{k_t} \frac{\as(k_t)}{2\pi}\,.
\end{equation}
This function $\cS(t)$ has been calculated numerically in the large
$\NC$ limit in \cite{dassalNG1} and progress in understanding the
dynamics associated with its all-orders form has been made in
\cite{dassalNG2,BanfiEtAlNG}.
\vspace{-0.3cm}

\subsection{Discontinuously global observables}
\label{sec:discontglobal}

\FIGURE{
    \epsfig{file=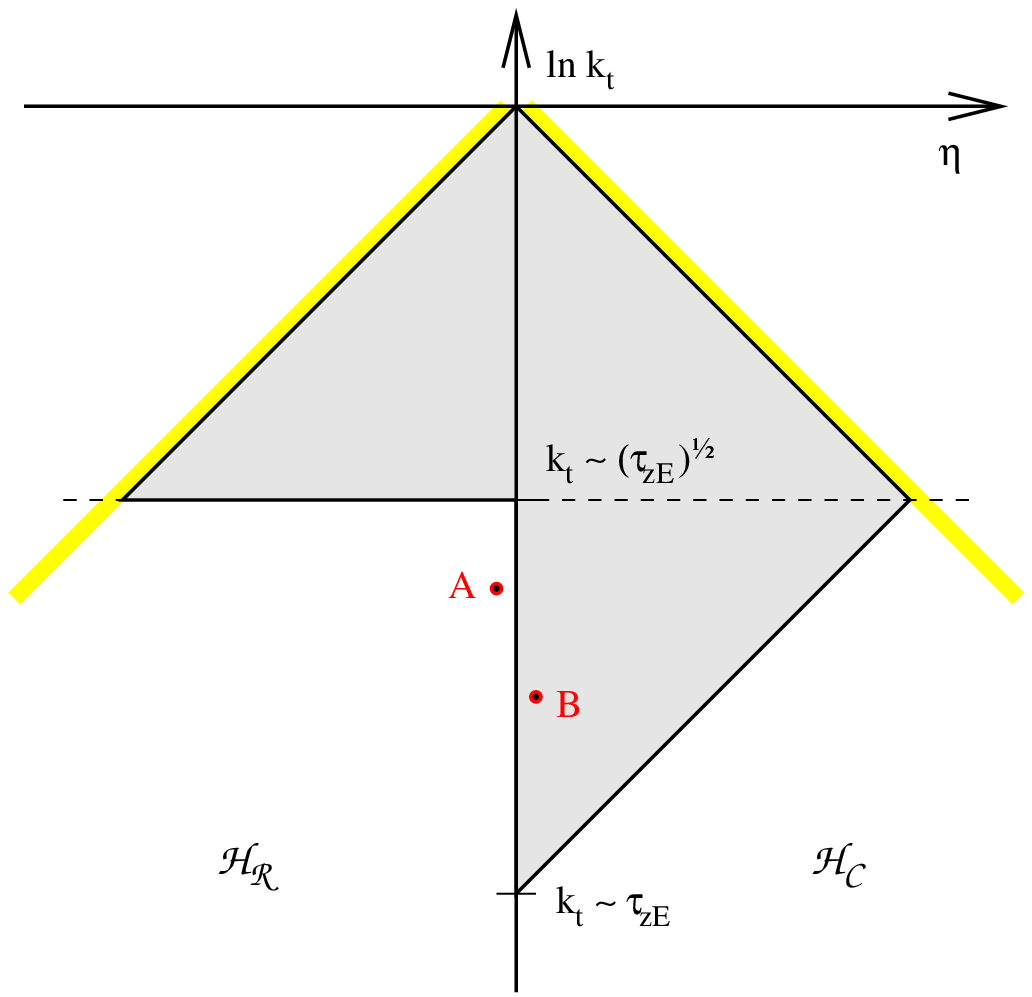,width=0.485\textwidth}
    \caption{Analogue of fig.~\ref{fig:discont}, shown for a
      discontinuously global observable, the photon-axis thrust,
      normalised to the energy in $\hc$, $\tau_{zE}$.}
    \label{fig:discontTze}
}

While the definition given above for a non-global observable is
sufficient to ensure the presence of non-global logarithms, the
converse is not true: it turns out that there exist observables that
are global, but which are nonetheless subject to non-global
logarithms.

An example is $\tau_{zE}$ whose value is 
\begin{equation}
  \tau_{zE} \simeq 2\left|\sum_i {\vec k}_{t,i}\right|^2 + 
  2\sum_i k_{t,i}e^{-|\eta_i|} \Theta(\eta_i)\,.
\end{equation}
The region of phase space to be excluded for the observable to have
smaller than some given small threshold is shown in
fig.~\ref{fig:discontTze}. The whole angular extent of phase space is
relevant, so the observable is global. However the dependence of the
observable on emissions in $\hr$ and $\hc$ is parametrically
different, going as $k_t^2$ and $k_t$ respectively. To denote this
kind of observable, we introduce a new term: a discontinuously global
observable. This is defined as being an observable, which, while
sensitive to emissions in the whole angular reach of phase space, has
a parametric dependence on $k_t$ (or energy) which changes
discontinuously across one or more boundaries in
angle.\footnote{Strictly speaking the parametric dependence may
  actually change smoothly, as long as the regions of different
  parametric dependence are separated by a limited region in angle.}

In the case of $\tau_{zE}$, non-global logarithms arise through
emission patterns of the kind represented by gluons A and B in
fig.~\ref{fig:discontTze}. Large-angle emission of $A$ into $\hr$ will
not contribute significantly to the observable as long as $k_{t,A} \ll
\sqrt{\tau}$. However it will modify the pattern of large-angle
emissions B in $\hc$ when $k_{t,B} \ll k_{t,A}$. On the other hand
large-angle emission B in $\hc$ must
be forbidden if $k_{t,B} \gtrsim \tau$. Therefore there is an
energy-ordered region $\sqrt{\tau} \gg k_{t,A} \gg k_{t,B} \gtrsim
\tau$ in which we should account for the modification of the emission
pattern of $B$ by the presence of $A$. This gives a non-global log,
which at second order in $\as$ contributes as
\begin{equation}
  \label{eq:twogluonDiscont}
   -\cf\ca \frac{\pi^2}{3} \left( \asb \int_\tau^{\sqrt{\tau}}
    \frac{dk_t}{k_t} \right)^2\,,
\end{equation}
\ie one quarter of the contribution in the fully non-global
$\tau_{tE}$ case, eq.~\eqref{eq:twogluonNG}. Analogously, at all orders
we will have the same function $\cS(t)$ as for the non-global
$\tau_{tE}$, but with a different integration range in the definition
of $t$:
\begin{equation}
  \label{eq:tdefDiscont}
  t = \int_\tau^{\sqrt{\tau}} \frac{d k_t}{k_t} \frac{\as(k_t)}{2\pi}\,.
\end{equation}
This non-global contribution was erroneously omitted in the original
resummation for $\tau_{zE}$ \cite{ADS}. This ought, in principle, 
to have been
revealed by the comparison with fixed-order calculations, however in
practice the numerical accuracy of the fixed-order calculations is
such that it is not possible to distinguish between comparisons with
and without the non-global logs (this is specific to $\tau_{zE}$ ---
other observables such as $\tau_{tE}$ have effects which are four
times larger and which therefore can be more straightforwardly
distinguished in comparisons to fixed-order calculations).

\subsection{Dynamically discontinuously global observables}

An even more subtle situation involving non-global logarithms is that
of the rather clumsily named `dynamically discontinuously global
observables.' These are observables which in a two-emission analysis
(such as figs.~\ref{fig:discont}, \ref{fig:discontTze}) appear to be
global without discontinuities, but which develop a discontinuity
dynamically at all orders. To illustrate what we mean, we consider the
broadening with respect to the photon axis, $B_{zE}$. This is given by
\begin{equation}
  \label{eq:brddef}
  B_{zE} \simeq |\sum_i {\vec k}_{t,i}| + \sum_i k_{t,i} \Theta(\eta_i)\,.
\end{equation}
 Note that the parametric dependence on $k_t$ is the same in both hemispheres
(there is a change in normalisation dependence in going from one
hemisphere to another, but this is not sufficient to induce
single-logarithms).

There is a subtlety however in that it is possible for $|\sum_i {\vec
  k}_{t,i}|$ to be parametrically smaller than some of the individual
$k_{t,i}$'s in $\hr$, \emph{if} there is a cancellation in the vector sum. So
for emissions in $\hr$ it is in principle possible to have $k_{t,i}
\gg B_{zE}$ and one is then faced with the issue that such emissions,
 when they are at large angles, modify the pattern of
emissions into $\hc$. At first sight this may seem quite
unlikely because of the small probability of a sufficient cancellation
in the vector sum. However as discussed in \cite{dassalbroad}, at
sufficiently small values of the broadening the favoured mechanism
for obtaining a small value of $B_{zE}$ 
is the cancellation in the vector sum,
rather than the suppression of emissions. In such a situation
non-global logarithms will arise.

In \cite{dassalbroad} we addressed the issue of this vector
cancellation in some detail, showing that it was associated with a
divergence in
the pure NLL resummation and discussing how to improve the resummation
to maintain the accuracy of there being only relative $\order{\as}$
corrections in the $\cR' \sim 1$ region. However that treatment is
incomplete since it did not take into account the non-global
logarithms that arise in this context due to the dynamically
discontinuous globalness.

From a phenomenological point of view, it is perhaps fortunate that
these issues become relevant only considerably to the left of the
Sudakov peak, a region that we ignore in the comparisons to data shown
later.  Indeed in the fits to data we use the pure NLL resummed
formula for the broadening, for which there are no non-global
logarithms.

\subsection{Phenomenological impact of non-global logs}

\FIGURE{
\epsfig{file=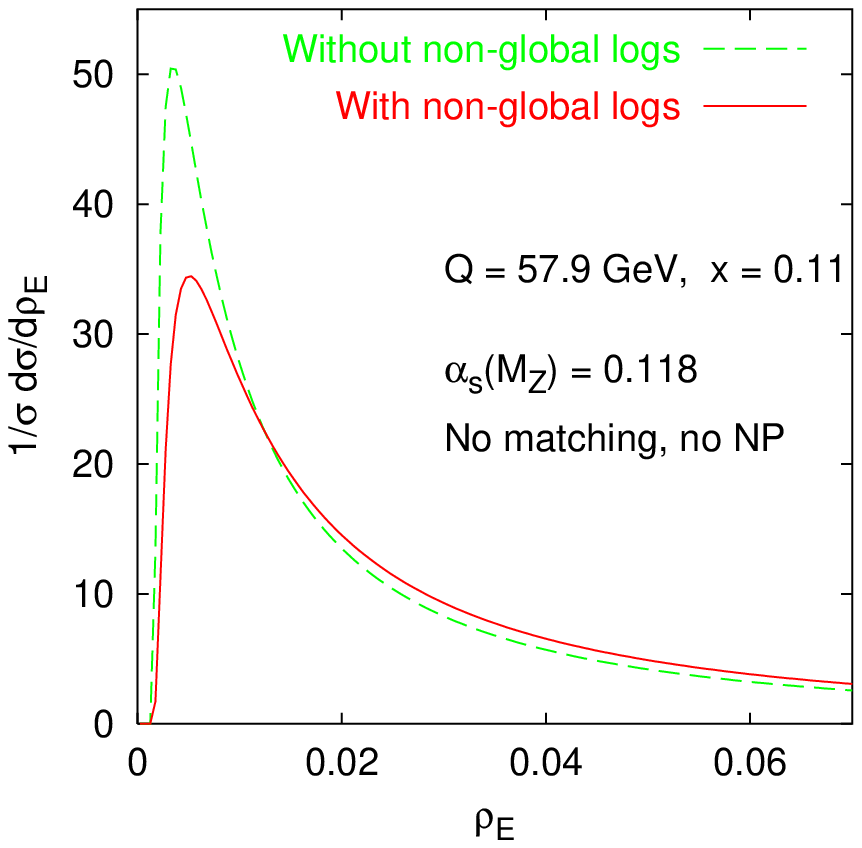,width=0.5\textwidth}
\caption{Resummation
  with and without non-global logs (shown without matching to fixed
  order).\label{fig:NGeffect}}}

In figure~\ref{fig:NGeffect} we illustrate the effect of non-global
logarithms in the resummation of $\rho_E$ (other non-global
observables are similar).  The figure compares (unmatched)
resummations with and without account of non-global logs. One sees
that neglecting them leads to an overestimate of the peak-height by
almost $50\%$, though elsewhere in the distribution, the effect is
smaller.  In practice, matching to fixed-order calculations reduces the
impact of neglecting the non-global logarithms, but the effects are
still of the order of $20$ to $30\%$.

The effect of non-global logs in the discontinuously global observable
$\tau_{zE}$ (not shown), is found to be smaller, at about the $10\%$
level. This is consistent with the naive expectation (\cnf
section~\ref{sec:discontglobal}) of a reduction by a factor of four
compared to the non-global case. In section~\ref{sec:data} we examine
the effect of the non-global logarithms in fits for $\as$ and the
non-perturbative $\alpha_0$ parameter.

\section{Singularities other than at $\boldsymbol{V=0}$}
\label{sec:nonsmooth}

The main aim of this paper is to discuss the resummation of the
singular behaviour of distributions around $V=0$. However, as was
pointed out in \cite{CataniWebberInternalDiv} there can also be
singular behaviour inside the phase space of
certain observables, the
notable example in $\ee$ scattering being the $C$-parameter.  In DIS
the problem turns out to be much more common than in $\ee$, even
involving (in some cases) two divergent structures inside the phase
space of a single observable. This is in part because one only counts 
particles in $\hc$ and hence there is a sharp angular cut-off (the interface 
of $\hc$ and $\hr$) which can introduce problems of non-smoothness in
the various event shapes.

Divergences inside the phase space are associated with discontinuous
or non-smooth structures at some lower order of perturbation theory.
Higher order corrections act to smooth out these features, but order
by order are divergent. One example of this is that in $\ee$ the LO
distribution for the $C$-parameter has a step function at $C=3/4$; the
inclusion of all higher orders smooths out the step to give a
`Sudakov shoulder,' $\sim \as e^{-\as \ln^2(C-3/4)}$
\cite{CataniWebberInternalDiv}, though order by order the shoulder is
associated with (a divergent series of singular) $\as^{n+1} \ln^{2n} (C-3/4)$
terms.

Such resummations, while interesting, are beyond the scope of this
article. One should nevertheless be aware of the kinds of
non-smoothness that arise in the different observables.  We shall
discuss in particular the DIS $C_E$-parameter, and then give a summary
of the properties of other observables.



\subsection{The $\boldsymbol{C}$-parameter}
It is convenient here to use the variables $z$ and $\xi$ defined in
\cite{DasWeb}, such that the incoming quark, outgoing
quark and outgoing gluon have respectively (Breit-frame) momenta $p$,
$r$ and $k$, where
\begin{subequations}
\begin{align}
  p &= \frac{Q}{2} \left(\frac1\xi,0,0,-\frac1\xi\right)\,,\\
  r &= \frac{Q}{2} \left(z_0,z_\perp,0,z_3\right)\,,\\
  k &= \frac{Q}{2} \left({\bar z}_0,-z_\perp,0,{\bar z}_3\right)\,,
\end{align}
\end{subequations}
with 
\begin{align}
  z_0 &= 2z-1 + \frac{1-z}{\xi}\,, \qquad & z_3 = 1 - \frac{1-z}{\xi},
  \nonumber\\
  {\bar z}_0 &= 1-2z + \frac{z}{\xi}\,, \qquad & {\bar z}_3 = 1 -
  \frac{z}{\xi},\quad\quad\!
  \nonumber\\
  z_\perp &= 2\sqrt{z(1-z)(1-\xi)/\xi}\,.
\end{align}
\begin{figure}[t]
  \begin{center}
\epsfig{file=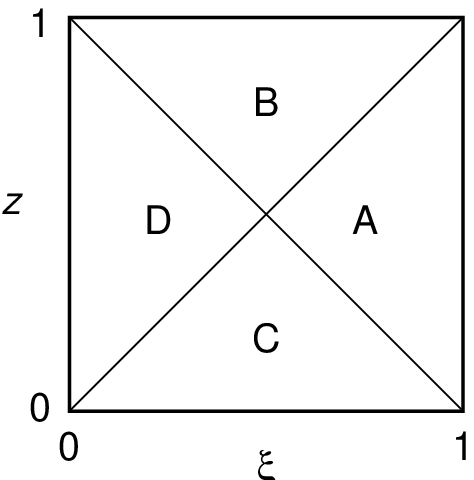,width=0.32\textwidth}\hfill
    \epsfig{file=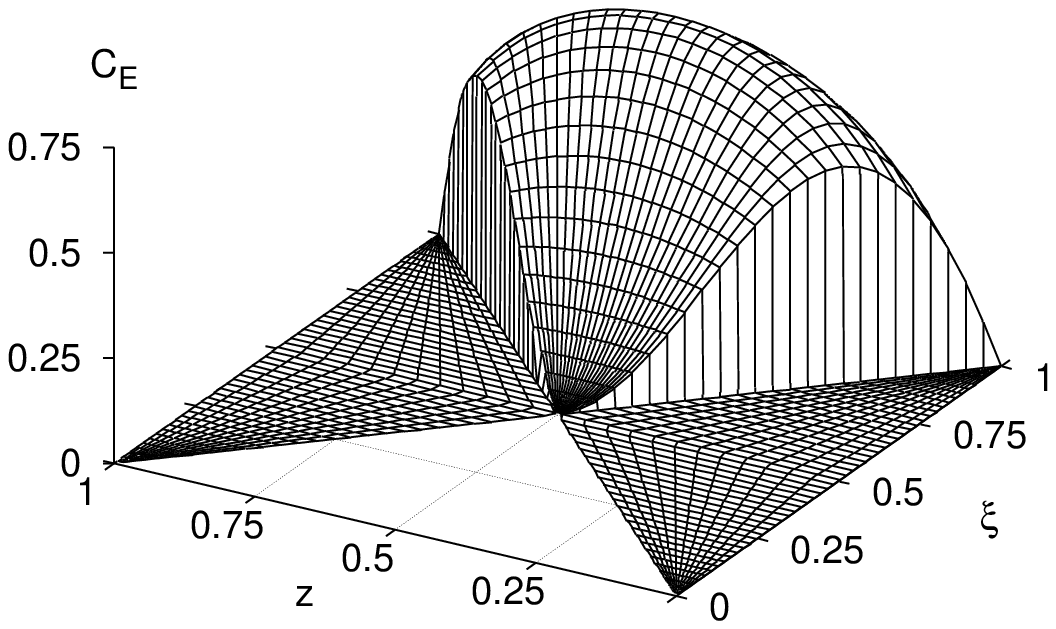,width=0.6\textwidth}
    \label{fig:Cprof}
    \caption{left, the four quadrants in the $\xi, z$ place; right,
      the value of $C_E$ as a function of $\xi$ and $z$.}
  \end{center}
\end{figure}%
The left-hand part of fig.~\ref{fig:Cprof} shows the different regions
in the $\xi,z$ plane:

\begin{center}
\begin{tabular}{lcl}
  A: $r$ and $k$ $\in \hc$ && B: $r \in \hc$\\
  C: $k \in \hc$ && D: $\hc$ empty
\end{tabular}
\end{center}
The $C_E$-parameter is undefined in region D, zero in regions B and C,
and in region A has the value \cite{DasWeb}:
\begin{equation}
  \label{eq:Cparamvalue}
  C_E = \frac{3(2\xi -1)^2 z_\perp^2}{z_0 {\bar z}_0}\,,
\end{equation}
as represented in the right-hand part of fig.~\ref{fig:Cprof}. Of
interest from the point of view of discontinuities of the distribution
and its derivatives are extrema (maxima) of \eqref{eq:Cparamvalue}.

For example $C_E$ takes its maximum value of $3/4$ at
\begin{equation}
  \xi_m = \frac12 \left(1 + \frac1{\sqrt{2}}\right)\,, \qquad z_m = \frac12\,.
\end{equation}
In the neighbourhood of this point
\begin{equation}
  C_E = C_{\mathrm{max}} + \frac12 c_{\xi\xi} (\xi - \xi_m)^2 +
  \frac12 c_{zz} (z - z_m)^2 + \cdots
\end{equation}
with the second derivatives $c_{\xi\xi}$ and $c_{z z}$ both
negative. The resulting distribution of $C_E$ in the neighbourhood of
$C_E = C_{\mathrm{max}}$ is a step function,
\begin{equation}
  \frac{d\sigma}{d {C_E}} \simeq  \Theta(C_\mathrm{max} - C_E)
  \left.\frac{d^2\sigma}{d\xi dz}\right|_{\xi_m,z_m}
  \frac{2\pi}{\sqrt{c_{\xi\xi}c_{zz}}}\,,
\end{equation}
in close analogy with the $\ee$ case
\cite{CataniWebberInternalDiv}. We note that if the maximum is not
smooth, but instead sharply peaked (\eg like the summit of a tetrahedron),
then the distribution goes smoothly to zero at the maximum, because
the measure associated with a given value of $V$ goes to zero linearly
with $V_\mathrm{max} - V$.

A second kind of feature that gives non-smooth distributions is a
smooth extremum along an edge, for example at
\begin{equation}
  \xi_c = \frac12 + \frac12 \sqrt{\sqrt{2}-1}\,, \qquad z_c = \xi_c\,, \;
  1-\xi_c
  \quad \Rightarrow \quad C_c = 3\left(3 - 2\sqrt{2}\right)\simeq 0.5147\,,
\end{equation}
along the borders of regions A and B and of A and C in
fig.~\ref{fig:Cprof}. To illustrate the effect this has on the
distribution let us write
\begin{equation}
  \label{eq:edgeExpansion}
  C = \Theta(x_\perp) \left(C_c + c_{x_\perp}
  \delta x_\perp + \frac12 c_{x_\prll x_\prll} \delta x_\prll^2 \right)\,,
\end{equation}
where $\delta x_\perp$ and $\delta x_\prll$ represent distances away
from the maximum in directions perpendicular and parallel to the edge
respectively.\footnote{We ignore the possibly non-zero $c_{x_\prll
    x_\perp}$ second derivative as it would not change the qualitative
  properties of the results.} In the case of the $C_E$-parameter,
$c_{x_\perp} > 0$, while $c_{x_\prll x_\prll} < 0$ and 
\begin{equation}
  \frac{d\sigma }{d C_E} \simeq 2
  \int_{\mathrm{max}\left(0,\frac{C_E-C_c}{c_{x_\perp}}\right)} 
  \frac{dx_\perp}{\sqrt{2|c_{x_\prll x_\prll}|(c_{x_\perp}{x_\perp} - (C_E-C_c))}}
  \left.\frac{d^2\sigma}{d{x_\perp} d{x_\prll}}\right|_{x_\prll =
    \sqrt{\frac{2(c_{x_\perp} {x_\perp} - (C_E-C_c))}{|c_{x_\prll x_\prll}|}}}\,,
\end{equation}
where we have already performed the integration over $x_\prll$.
The easily observed fact that the $x_\perp$ integral is not dominated by the
small $x_\perp$ region suggests that it is not sufficient to be
considering only the region close to the edge-maximum in
\eqref{eq:edgeExpansion}.  However for $C_E < C_c$, the qualitative
behaviour of $d\sigma /dC_E$ can nevertheless be understood by
ignoring the $x_\prll$ dependence of $\frac{d^2\sigma}{d{x_\perp}
  d{x_\prll}}$:
\begin{align}
  \frac{d\sigma }{d C_E} &- \left.\frac{d\sigma }{d C_E}\right|_{C_c} 
  \nonumber \simeq\\
  &\simeq
  2\left.\frac{d^2\sigma}{d{x_\perp} d{x_\prll}}\right|_{x_\perp = x_\prll =0}
  \int_0^\infty dx_\perp \frac{1}{\sqrt{2|c_{x_\prll x_\prll}|}}
  \left(\frac{1}{\sqrt{(c_{x_\perp} {x_\perp} - (C_E-C_c))}}
    - \frac{1}{\sqrt{c_{x_\perp} {x_\perp}}}\right) \nonumber\\
  &\simeq 
    -\left.\frac{d^2\sigma}{d{x_\perp} d{x_\prll}}\right|_{x_\perp = x_\prll =0}
    \frac2{c_{x_\perp}}\sqrt{\frac{2(C_c - C_E)}{|c_{x_\prll
          x_\prll}|}}\,,
    \qquad\quad(C_E < C_c)\,.
\end{align}
In other words below $C_c$ there is a rapid fall of the distribution,
behaving as a constant minus a term proportional to $\sqrt{C_c - C_E}$
just below the critical point. When approaching the critical point
from above rather than from below the distribution remains smooth.

\begin{figure}[t]
  \begin{center}
    \epsfig{file=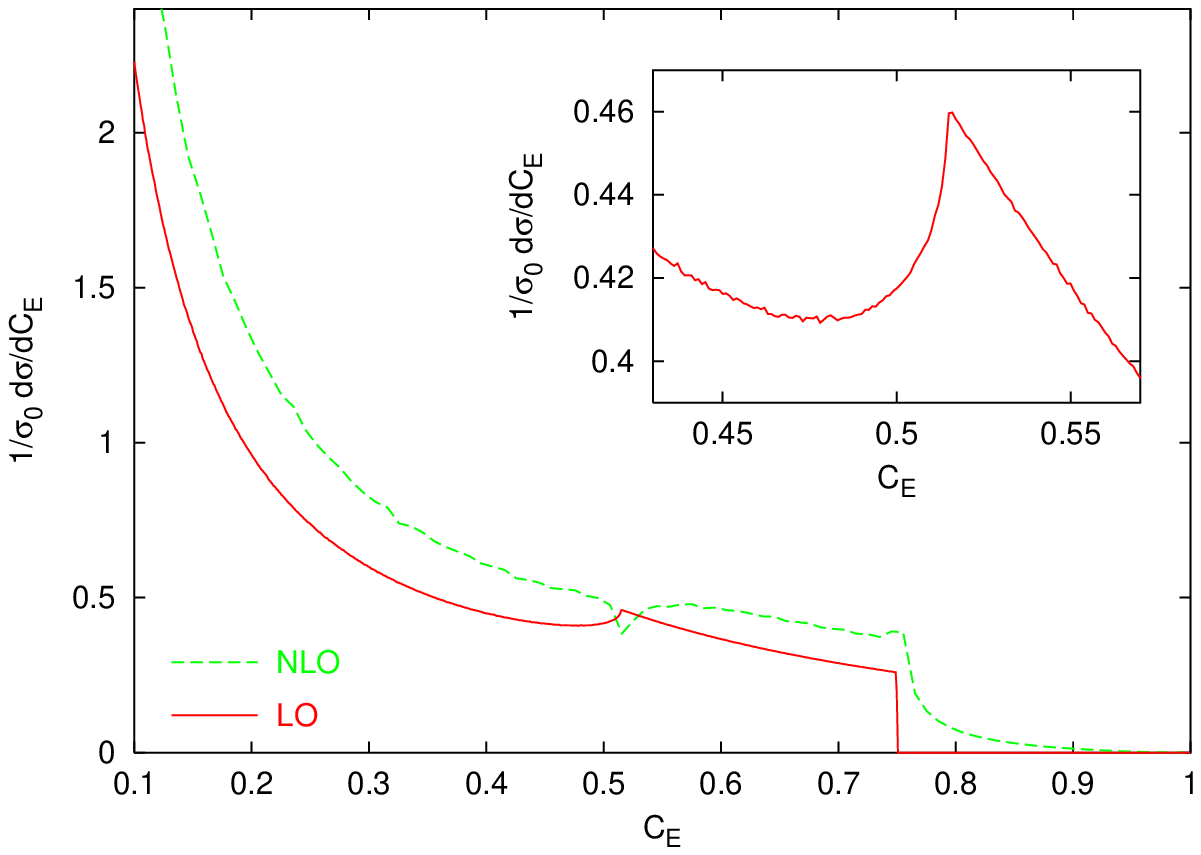,width=0.7\textwidth}
    \caption{Distribution of the $C_E$ parameter for $Q=57.9$~GeV and
      $x=0.11$ at LO and NLO (both normalised to the Born cross
      section). Both contributions have been calculated with \disent
      \cite{DT}.}
    \label{fig:CdistSteps}
  \end{center}
\end{figure}

Both the step function at $C_\mathrm{max}$ and the square-root
behaviour at $C_c$ can be seen in the leading-order distribution for
the $C_E$-parameter in fig.~\ref{fig:CdistSteps}. The figure shows
also the LO$+$NLO result. The pathological nature of the structures
which arise at NLO is only partially evident because of the limited
resolution of the NLO fixed order terms. 

\subsection{Other observables}

The kinds of patterns that we have discussed can be relevant for other
observables --- i.e.\ a smooth (quadratic) maximum $V(\xi,z)$ in
$2$-dimensions leads to a step function in the distribution of $V$,
while a smooth maximum along an edge in $\xi,z$ leads to a square-root
behaviour for the distribution around that edge-maximum. The sign of
the coefficient of the square-root is the same as the sign of
$c_{x_\perp}$.

It should be noted that features such as step functions (especially at
the upper boundary of phase space) can also arise from other kinds of
maxima, in particular an edge of constant height (i.e.\ $c_{x_\prll}
=c_{x_\prll x_\prll} \equiv 0$ and $c_{x_\perp} < 0$). One such
example is the $\tau_{zE}$ observable whose distribution has a step
function at $\tau_{zE}=1$, which is the upper phase-space limit at all
orders. Note however that, at least at NLO, there seems to be no
higher-order `pathological' behaviour associated with this particular
discontinuity of the distribution, perhaps because of the different
origin of the discontinuity relative to those discussed above.

Given that these questions are not the main topic of this article, we
shall not enter into any more detail. We do however give a summary of
the non-smooth features of all the different observables discussed at
various stages of this article, table~\ref{tab:features}.

\begin{table}[htbp]
\begin{center}
\begin{tabular}{|c|l|l|}
  \hline
  Var & Feature & Position \\ \hline
  $C_E$     & $\Theta(C_\mathrm{max} - C_E)$            &$C_\mathrm{max}=
  3/4$ (LO upper limit)     \\ 
            & $\mathrm{const.} -\sqrt{C_c - C_E}$ 
            & $C_c = 3(\sqrt{2}-1)^2\simeq 0.5147$ \\ \hline
  $\tau_{tE}$ & $\sqrt{\tau_\mathrm{max} - \tau_{tE}}$  
  & $\tau_\mathrm{max} \simeq 0.293$ (LO upper limit)    \\  
      & $\mathrm{const.} -\sqrt{\tau_c - \tau_{tE}}$ & $\tau_c \simeq 0.134$ \\ \hline
  $\tau_{zE}$ & $\Theta(\tau_{\mathrm{max}} - \tau_{zE})$  &
  $\tau_{\mathrm{max}} = 1$ (all-orders upper limit) \\
  & & [NB: NLO seems well-behaved] \\ \hline
  $\rho_{E}$ & $\Theta(\rho_\mathrm{max} - \rho_{E})$ &
  $\rho_\mathrm{max} = 1/4$ (all-orders upper limit) \\ \hline
  $B_{zE}$ & $\Theta(B_\mathrm{max} - B_{zE})$  & $B_\mathrm{max} =
  1/2$ (all-orders upper limit) \\ \hline 
           
\end{tabular}
\end{center}
\caption{List of the non-smooth `features' of a range of DIS
  observables at LO. Except where stated, they appear to be associated
  with divergent (in $V$) or, at the very least, perturbatively poorly
  convergent structures at higher order.}
\label{tab:features}
\end{table}

\section{Speeding-up fixed-order calculations}
\label{sec:speedyMC}

A necessary element in the comparison of our predictions to data is
the matching of the resummed results with fixed-order
calculations. However the determination of high-resolution fixed order
predictions with a reasonable precision is a significant technical
obstacle. High resolution is needed,
even when data are coarsely binned, because the shift (or convolution)
of the distribution coming from non-perturbative corrections will
displace the effective position of the bin boundaries in the
perturbative distribution by an a priori unknown amount.

Four programs exist for NLO final-state
calculations in DIS: \textsc{mepjet} \cite{MEPJET}, \disent \cite{DT},
\disaster \cite{DR} and \textsc{nlojet} \cite{NLOJET} (not available
when this project was started). \textsc{mepjet} has been found
\cite{DISNLOComparisons} to give substantially different results from
the other programs (this is unfortunate, because it is the only one
with the option of calculating $Z$-exchange contributions) and
additionally is apparently not suited to the determination of event-shape
distributions.

Accordingly at the start of this project, \disent and \disaster were
chosen for generating fixed-order distributions. For certain
observables however, non-negligible differences had been observed
between their predictions \cite{DISNLOComparisons}. Comparisons to
resummations of $\tau_{zQ}$ established \cite{ADS} that some
logarithmically enhanced terms ($\as^2 L^3$ in the integrated shape
cross-section) were incorrect in \disent and the problem has yet to be
fixed. This formally leaves us with only \disaster as an option.

%
%
%
But \disaster is slow. On a
computer with a 2~GHz Pentium~IV processor,\footnote{Corresponding to
  SPEC CFP2000 of about 700 \cite{SPEC}.} two billion
events\footnote{Roughly what is used in the coming sections --- in
  finely-binned distributions, statistical errors remain visible, \eg
  as for the NLO distribution for the $C$-parameter in
  fig.~\ref{fig:CdistSteps}.}
represents 50 days' computing time. Within the standard approach for
such programs, one does a run at each $x,Q$ pair of experimental
interest. The resulting requirement of 400 days' computing time would
have been prohibitive, especially given that at the time when the
calculations were carried out, the computers available to us were a
factor of two or three slower than those currently on the market.

An alternative approach is based on the observation that for fixed
Bjorken-$y$, all that changes when going to different $x$ and $Q$
values is the parton distributions. The matrix element and event-shape
calculations remain the same, and there is no need to repeat them. In
other words one can exploit factorisation to separate out the
structure function ($x,Q$) dependence from the rest of the problem. An
algorithm which does this (accounting also for Bjorken-$y$ dependence)
is presented below. It has been implemented in a wrapper program,
\dispatch, which is available from
\texttt{http://www.lpthe.jussieu.fr/\~~\!\!\!salam/dispatch/}. In
practice, with \disaster it allowed us to calculate distributions at $8$
$x,Q$ points using only about $30\%$ more computing time than would
have been needed for a single $x,Q$ point. With \disent the overhead
of processing several $x$ and $Q$ points is larger in relative terms
(because the matrix-element calculations are quicker in
\disent$\!\!$), but we still need only about $2.5$ times more
computing time for $8$ $x,Q$ points than for a single one.

\subsection{The dispatch algorithm}

The algorithm, as currently implemented, is suitable for any
observable whose definition, when expressed in the Breit frame,
contains no dependence on $x$ or on any dimensionful quantity other
than through its ratio with $Q$ (so for example there must be no
dependence on the incoming proton momentum, or on the proton remnant,
nor any cuts on absolute momenta, nor cuts in the lab frame).
Breit-frame event shapes are a good example of such observables.

We shall present our discussion in terms of weighted cross sections
$\sigma^{(w)}$, where the weighting function will for example be the
value observable itself if one wants to calculate its mean value, or
a product of $\Theta$ functions so as to obtain the cross
section for the observable's value to be in a certain `bin'. For such
weighted cross sections we can write
\begin{equation}
  \frac{x Q^4 d\sigma^{(w)}}{dx dQ^2} = 2\pi
    \alpha_\mathrm{em}^2
  \left((1+(1-y)^2)F^{(w)}_2 - y^2 F^{(w)}_L \right) \equiv
  \cD^{(w)}(x,y,Q^2)\,, \qquad y = \frac{Q^2}{xs}\,, 
\end{equation}
where we have introduced a shorthand $\cD^{(w)}(x,y,Q^2)$ for the weighted
differential cross section. As is standard for the total cross section
we have decomposed the result into transverse+longitudinal ($2$) and
pure longitudinal ($L$) weighted structure functions,
$F^{(w)}_{2,L}$.  These two weighted structure functions are
themselves convolutions of weighted coefficient functions with parton
distributions (both of which are vectors in flavour space):
\begin{equation}
  F^{(w)}_{2,L}(x,Q^2) = \int_x^1 \frac{d\xi}{\xi}\,
  \bC^{(w)}_{2,L}(\xi, \as)
  \cdot x \bq\left(\frac{x}{\xi},Q^2 \right)\,.
\end{equation}

A Monte Carlo program such as \disent or \disaster returns event
contributions which, after taking into account the weighting function
and averaging over many events with the same value of $x/\xi$,
correspond to the following function (again a vector in flavour space)
\begin{equation}
  \label{eq:bcD}
  \bcD_n^{(w)}(\xi,y) = 2\pi \alpha_\mathrm{em}^2 \left(
    (1+(1-y)^2) \bC^{(w)}_{2n}(\xi) - y^2 \bC^{(w)}_{Ln}(\xi)\right)\,.
\end{equation}
where the $n$ refers to the order of perturbation theory. There are
two points of interest. Firstly, order by order, $\bcD_n^{(w)}$ is a
function only of $\xi$ and $y$, but not of $x$ or $Q^2$. Therefore, for
a fixed $y$ value, the event contributions provided by the Monte Carlo
at a single $x$ and $Q^2$ value are sufficient to allow us to
calculate $\cD_n(x,y,Q^2)$ for any $x$ and $Q^2$ value, through the
following convolution
\begin{equation}
  \cD_n^{(w)}(x,y,Q^2) = \int_x^1 \frac{d\xi}{\xi} \, 
  \bcD_n^{(w)}(\xi,y) \cdot x
  \bq\left(\frac{x}{\xi},Q^2 \right)\,.
\end{equation}
Accordingly the \dispatch\ program uses Monte Carlo events at a single
value of $x$ and $Q^2$ (chosen automatically) to calculate the
$\cD_n^{(w)}$ for a set of $x$ and $Q^2$ values, all with the same
value of $y$.

For comparison to data, a fixed value of $y$ is usually not suitable.
However the second point of interest in eq.~\eqref{eq:bcD} is that all
the $y$ dependence comes from the linear weighting of the two
$y$-independent functions, $\bC_{2n}$ and $\bC_{Ln}$. So if we have
calculated $\cD_n^{(w)}$ for two values of $y$, say $y_1$ and $y_2$,
then we can obtain it for any third value, $y_3$, as follows:
\begin{multline}
  \cD_n^{(w)}(x,y_3,Q^2) = \frac{(y_2y_3 - y_2 - y_3)(y_3-y_2)}{
    (y_1y_2 - y_1 - y_2)(y_1-y_2)}\cD_n^{(w)}(x,y_1,Q^2) \\
  - \frac{(y_1y_3 - y_1 - y_3)(y_3-y_1)}{
    (y_1y_2 - y_1 - y_2)(y_1-y_2)}\cD_n^{(w)}(x,y_2,Q^2)\,.
\end{multline}
So one simply carries out two \dispatch\ runs at different $y$ values
and then uses a utility program to combine them with the correct
weights so as to correspond to a fixed value of $ep$ centre of mass
energy, $\sqrt{s}$.

This means that in principle we have to do two full runs of \dispatch.
However in many practical situations, for example when comparing to
the H1 event-shape distributions in \cite{H1NewData} the relevant $y$
values all satisfy $y \lesssim 0.4$ --- choosing a low value for $y_1$
and a high value for $y_2$ it turns out that the weight for the
high-$y$ run is usually somewhat smaller than that for the low-$y$
run, allowing one to generate fewer events at the high-$y$ point. As
an example, for the fixed-order distributions needed in this paper, 
we used $y_1=10^{-3}$ and $y_2=0.9$,
and generated only one tenth of the total number of Monte Carlo events
at the higher $y$ value. So there is relatively little overhead from
generating the second high-$y$ point.

\section{Inclusion of non-perturbative effects}
\label{sec:NP}

All of the observables that we consider in this article receive
significant non-perturbative contributions. So before comparing our
perturbative resummed results to data, it is necessary to correct them
for non-perturbative effects. Until a few years ago, this could only
be done using the difference between parton and hadron levels in Monte
Carlo event generators such as Herwig \cite{Herwig} or Pythia
\cite{Pythia}. However such procedures tend to be ill-defined, because
of ambiguities in the separation between perturbative and
non-perturbative effects. Furthermore, though very successful in
describing the data, they involve a number of parameters that must be
tuned to the data, as well as many model-dependent assumptions.

When studying event shapes, it can be argued that many of the details
of hadronisation (hadron multiplicities, relative abundances of
different hadron species, etc.) factor out from the problem, due to
the semi-inclusive nature of the observable. This makes it possible to
consider the more inclusive aspects of hadronisation analytically,
within much simpler and more transparent theoretical approaches.

A general conclusion from such studies is that the dominant
hadronisation correction to event shapes is of order $\Lambda/Q$
\cite{ManoharWise,Webber94,DW,AK,BB,KS}. Furthermore there exists a
\emph{universality} hypothesis whereby the size of non-perturbative
corrections should involve the product of a calculable,
observable-dependent parameter (which we call $c_\cV$) and an a priori
unknown parameter (which we will later call $\alpha_0$) which should
be universal, \ie independent of the observable and of the process.
This hypothesis can be tested by fitting for $\alpha_0$ and $\as$ in a
range of observables and processes.\footnote{One may ask why one
  should be fitting both $\as$ and $\alpha_0$, given that $\as$ is
  already well determined from other measurements. Two answers exist:
  firstly one may actually wish to provide an alternative measurement
  of $\as$. Secondly any observable will be subject to higher-order
  perturbative corrections, and to fix $\as$ while fitting $\alpha_0$
  implies absorbing these higher-order
  corrections entirely into $\alpha_0$.} %
Substantial evidence in favour of universality has been obtained with
$\ee$ data, using both mean values and distributions
\cite{UniversalityTest} and also from mean values of DIS observables
\cite{H1NewData,ZEUS}.  In general distributions tend to give a much
more robust test of universality. This is because with mean values one
has one data-point for each $Q$ value and the fit distinguishes
$\alpha_0$ and $\as$ through their different $Q$ dependences. This
procedure can lead to strong correlations between results for $\as$
and $\alpha_0$ which can mask inconsistencies in the universality
hypothesis (or errors in the assumptions behind the calculation of the
coefficient $c_\cV$).  Distributions of event shapes on the other
hand, involve many data points at each $Q$ value and so have far
more discriminatory power.  This is one of the main motivations for
extending the DIS studies to distributions. It should be kept in mind
however that the extension to distributions is not completely without
problems, as we illustrate below.

\subsection{Means}

For mean values non-perturbative effects change the observable by an
amount $\langle \delta \cV \rangle$, which within the approach of
\cite{DW,Milan2} is assumed to be of the form
\begin{equation}
  \langle \delta \cV \rangle  = c_\cV \cP +
  \order{\frac{\as \Lambda}{Q}} + \order{\frac{\Lambda^2}{Q^2}}
\end{equation}
where
\begin{equation}
  \label{eq:cPfin}
 \cP\> \equiv\>  \frac{4C_F}{\pi^2}\cM \frac{\mu_I}{Q}
\left\{ \alpha_0(\mu_I)- \as(\mu_R)
  -\beta_0\frac{\as^2}{2\pi}\left(\ln\frac{\mu_R}{\mu_I} 
    +\frac{K}{\be_0}+1\right) \right\}\>,  \quad \as\equiv \al_{\MSbar}(Q)\>,
\end{equation}
with $\mu_I$ an arbitrary infrared matching scale (conventionally
taken to be $2$~GeV) and $\cM \simeq 1.49$ the Milan factor
\cite{Milan1,Milan2,DasWebMi,AnlMilan}. In this approach $\alpha_0$ is
interpreted as being the average value of an infrared finite
$\as$ between scales $0$ and $\mu_I$. The perturbative
subtraction terms eliminate the double counting between perturbative
and non-perturbative contributions (or equivalently they cancel the
perturbative renormalon contribution) and should ensure that the
resulting correction is independent of the infrared scale $\mu_I$
(though $\alpha_0(\mu_I)$ itself is not). The renormalisation scale
($\mu_R$) dependence of the perturbative subtraction terms has the
interesting property that it tends to cancel part of the
renormalisation scale dependence of the perturbative answer.

\subsection{Distributions}
For distributions, the situation is more complicated because
non-perturbative effects can modify the whole shape of the
distribution --- this implies a dependence on more than just a single
non-perturbative parameter. A very general relation between a
perturbative distribution $D_{\mathrm{PT},\cV}(v)$, and the full
result including non-perturbative effects, $D_{\mathrm{full},\cV}$,
can be written as
\begin{equation}
  \label{eq:shapeGenConvoluion}
  D_{\mathrm{full},\cV}(v) = \int dx \,f_\cV(x, v, \as(Q), Q)\,
  D_{\mathrm{PT},\cV}\left(v - \frac{x}{Q}\right)\,.
\end{equation}
For suitable observables (strictly speaking those that are linear in
the momenta of several soft particles), it is argued that in the
$2$-jet region non-perturbative effects are independent of $v$ and of
$\as$ and can be encoded through a simpler \emph{shape function}
$f_\cV(x)$ \cite{KS,MoreKS},
\begin{equation}
  \label{eq:shapeConvoluion}
  D_{\mathrm{full},\cV}(v) \simeq \int dx \,f_\cV(x)\,
  D_{\mathrm{PT},\cV}\left(v - \frac{x}{Q}\right).
\end{equation}
One can rewrite this in terms of integer
moments $f_{\cV,n}$ of $f_\cV(x)$,
\begin{equation}
  \label{eq:momentExpansion}
  D_{\mathrm{full},\cV}(v) = D_{PT,\cV}(v) + 
  \sum_{n=1}^{\infty} (-1)^n \frac{f_{\cV,n}}{n! Q^n} D_{PT,\cV}^{(n)} (v)\,,
\end{equation}
where $D_{PT,\cV}^{(n)}$ is the $n^\mathrm{th}$ derivative of
$D_{PT,\cV}$. Consistency with the mean value implies $f_{\cV,1}/Q
\equiv c_\cV \cP$. 

The higher moments each involve further non-perturbative
unknowns. As discussed in \cite{KS,MoreKS,DGE} the fact that
they multiply $D_{PT,\cV}^{(n)}(v) \sim 1/v^n$ implies an expansion in
powers of $1/(vQ)$, which in the limit of small $v$ should be
resummed. In other words formally one needs complete
knowledge of the shape function. This requirement is the price one
has to pay to `gain access' to all the extra information in
distributions as compared to mean values.

It is of interest to consider to what extent it is truly necessary to
have complete knowledge of the shape function. Let us take a generic
resummation where the normalised integrated distribution is
\begin{equation}
  \Sigma_q(v) = \exp( - \cR(v))\,,
\end{equation}
(for simplicity we neglect all NLL terms) and define $\cR'(v) =
d\cR/d\ln(1/v)$. The resummation is valid down to $\as \ln 1/v =
\order{1}$, which translates to $\cR'(v) = \order{1}$.  The peak of
the distribution
\begin{equation}
  \frac{d\Sigma_q}{dv} = \frac{\cR'}{v} \exp( - \cR(v))\,,
\end{equation}
is at $\cR' = 1 + \order{\as}$, so let us take this as the point up to
which we wish to calculate the distribution. To understand the
hierarchy of the expansion \eqref{eq:momentExpansion} one needs to
know the value of $v$ at $\cR'=1$. This depends strongly on the
observable being considered, but can be written fairly generally as
\begin{equation}
  v(\cR'=1) \sim \left(\frac{\Lambda}{Q}\right)^{p}\,,
\end{equation}
where for $B_{zE}$ and the $\ee$ total and wide-jet broadenings,
\begin{subequations}
\begin{equation}
  p = \frac{1}{1 + \frac{2\CF}{\pi\beta_0}} \simeq 0.42\,,
\end{equation}
(numerical values are given for $\nf=5$); for the $\ee$ thrust,
$C$-parameter and heavy jet mass, and for $\tau_{zQ}$ it is
\begin{equation}
  p = \frac{e^{\frac{\pi\beta_0}{2\CF}} -
    1}{e^{\frac{\pi\beta_0}{2\CF}} - \frac{1}{2}} \simeq 0.68\,,
\end{equation}
and for $\tau_{tE}$, $\rho_E$ and $C_E$ it is
\begin{equation}
  p = \frac{e^{\frac{\pi\beta_0}{\CF}} -
    1}{e^{\frac{\pi\beta_0}{\CF}} - \frac{1}{2}} \simeq 0.87\,.
\end{equation}
For $\tau_{zE}$ there is no analytical solution, but numerically,
\begin{equation}
  p \simeq 0.80\,.
\end{equation}
\end{subequations}
So for the broadenings (for which a shape function is in any case not
sufficient to describe the non-perturbative effects) the neglect of
higher moments might be justifiable down to the peak region, whereas
for the observables derived in this paper this is less likely to be
true.

Physically, an interesting approximation is to replace the shape function
by a $\delta$-function, which just shifts the perturbative
distribution \cite{KS,DokWeb97},
\begin{equation}
  D_{\mathrm{shift},\cV}(v) =  D_{\mathrm{PT},\cV}\left(v -
    c_\cV \cP \right)\,.
\end{equation}
Remaining terms associated with the full shape function would then
correct $D_{\mathrm{shift},\cV}(v)$ as follows
\begin{equation}
  D_{\mathrm{full},\cV}(v) = D_{PT,\cV}(v) + 
  \sum_{n=2}^{\infty} (-1)^n \frac{f_{\cV,n} - (f_{\cV,1})^n}{n! Q^n}
  D_{\mathrm{shift},\cV}^{(n)} (v)\,, 
\end{equation}
While this does not improve the formal convergence of the expansion
compared to \eqref{eq:momentExpansion} (we still have an expansion in
powers of $1/vQ$) one expects partial cancellation between $f_{\cV,n}$
$(f_{\cV,1})^n$ to improve the actual convergence. Indeed for some
observables in $\ee$, perturbative arguments \cite{DGE} explicitly suggest
$f_{\cV,2}  = (f_{\cV,1})^2$ and this seems also to be supported by
the data.

The approximation of the shift is very powerful, because one has only
a single non-perturbative parameter $\alpha_0$, which is supposed to
be universal between different observables, different processes and
both mean values and distributions.\footnote{Universality has been
  postulated (and tested against data) between shape functions (or
  components thereof) for restricted sets of observables in
  \cite{KorchemskyTafat} and \cite{DGE}, though some of the
  phenomenological aspects of the former analysis have subsequently
  been argued to be problematic \cite{SalamWicke} due to a neglect of
  hadron mass effects.} %
So in what remains of this article we will restrict our analyses to a
shift.

\subsection{Other considerations}

\subsubsection{Broadenings}

It was stated above that the use of a $v$ and $\as$-independent shape
function held for linear observables. One particular case where this
is known not to be true is that of the the broadenings
\cite{eeBrdNP,dassalbroad} and analogous multi-jet observables
\cite{MilanMultiJet}. The non-linear dependence of the broadening on
multiple emissions has the consequence that the typical
non-perturbative correction is roughly proportional to $\ln 1/\theta$,
where $\theta$ is the angle of the current-quark with respect to the
photon (or thrust) axis. Because of a correlation between $\theta$ and
the broadening, one finds that a non-perturbative correction
proportional to $\ln 1/B$. This is taken into account in the fits that
follow.

\subsubsection{Hadronic mass effects}
\label{sec:masseffects}

When testing the universality of $1/Q$ corrections, care is needed in
the treatment 
of hadronic mass effects. Most observables are defined in terms of
$3$-momenta. Some however, in particular jet masses, depend in their
conventional definitions on differences between particle energies and
3-momenta. This leads to universality-breaking $\Lambda/Q$ corrections
(enhanced by powers of $\ln Q$) \cite{SalamWicke}. Such contributions
can be eliminated by suitable redefinitions of the observables. Here
we shall consider observables defined in the $p$-scheme, meaning that
in the definition of the observable all occurrences of a particle's energy
are replaced by the modulus of its 3-momentum. Of the DIS observables
considered in this article, the only one to be modified is the jet
mass. Phenomenologically the use of a consistent scheme has been found
to be of vital importance in analyses \cite{SalamWicke,DGE}
of $\ee$ measurements.

Strictly speaking in the $p$-scheme there remain small
universality-breaking mass effects, however they are in practice
expected to be negligible, and given that we do not dispose of data in
properly universal schemes such as the $E$-scheme or a decay scheme
(where measurements are `carried out' after the decay of all massive
particles) we make do with $p$-scheme measurements.

Of more concern perhaps is that fact that there can be universal mass
effects which could give logarithmic corrections to the $Q$-dependence
of power-suppressed effects. Because we are not able to estimate the
absolute normalisation of these universal mass effects (or even their
sign) it is not a priori clear how to eliminate them. However it
should be noted that even `traditional' $1/Q$ corrections may be
subject to anomalous dimension-like corrections, coming from
single-logs $(\as \ln Q/\Lambda)^n$ associated with the fact that soft
particles are emitted not off a bare $q\qbar$ current, but rather off
a `dressed' $q\qbar$ pair: \ie the $q\qbar$ emits large angle soft
gluons with $k_t$'s ranging between between $Q$ and $\Lambda$ (giving
powers of $\as \ln Q/\Lambda$) and soft gluons with transverse momenta
of order $\Lambda$ are emitted coherently off that whole ensemble. The
approximate rapidity-independence of the `dressing' ought to ensure
however that universality is not broken.

A final point relevant to all these anomalous-dimension-like
corrections (whether associated with mass effects or coherent emission
off dressed $q\qbar$ pairs) is that in distributions, the single logs
$(\as \ln Q/\Lambda)^n$ may acquire a dependence on the value of the
observable. For example, for a continuously-global observable such as
$\tau_{zQ}$ one schematically expects occurrences of $(\as(Q) \ln
Q/\Lambda)^n$ in corrections to the mean, to be replaced by $(\as(\tau
Q) \ln \tau Q/\Lambda)^n$ in distributions, because the hardest
(perturbative) large-angle emission must have a scale which is smaller
than $\order{\tau Q}$. For non-global and discontinuously global
observables, the modification is likely to be more complex.

In any case it should be emphasised that even the first order
correction term, $\as \ln Q/\Lambda$, has yet to be calculated for any
observable. So in the phenomenological analyses that follow later we
shall ignore these complications.

\section{Resummation: impact and uncertainties}
\label{sec:impact}


\subsection{Effect of resummation and matching}

\FIGURE{\epsfig{file=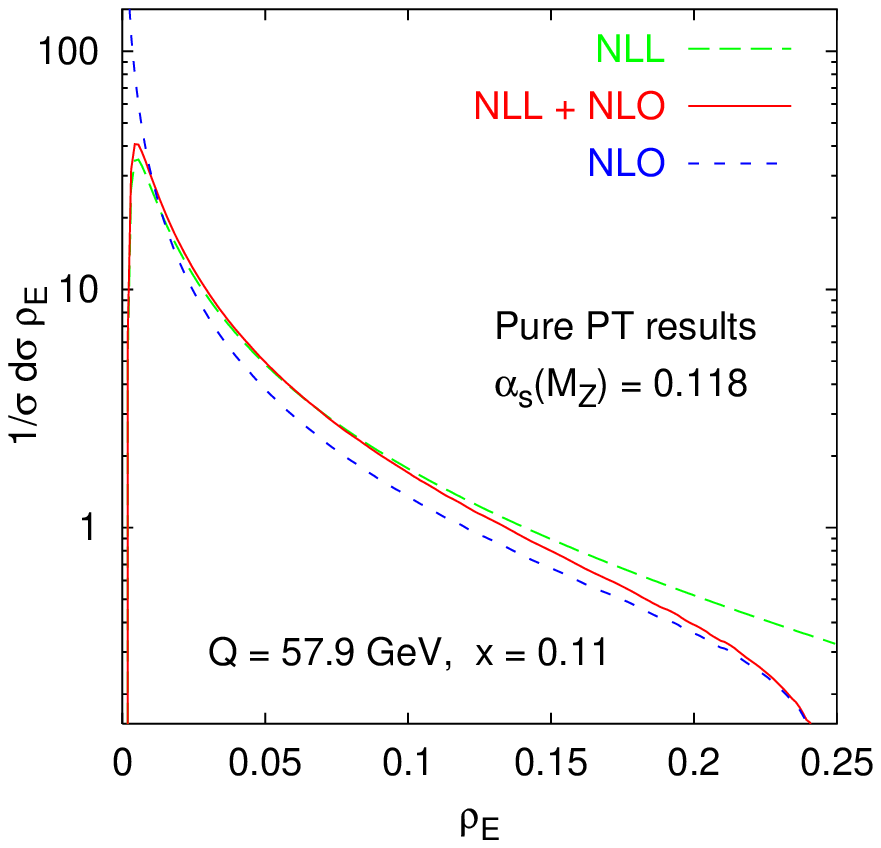,height=0.48\textwidth}\hfill\epsfig{file=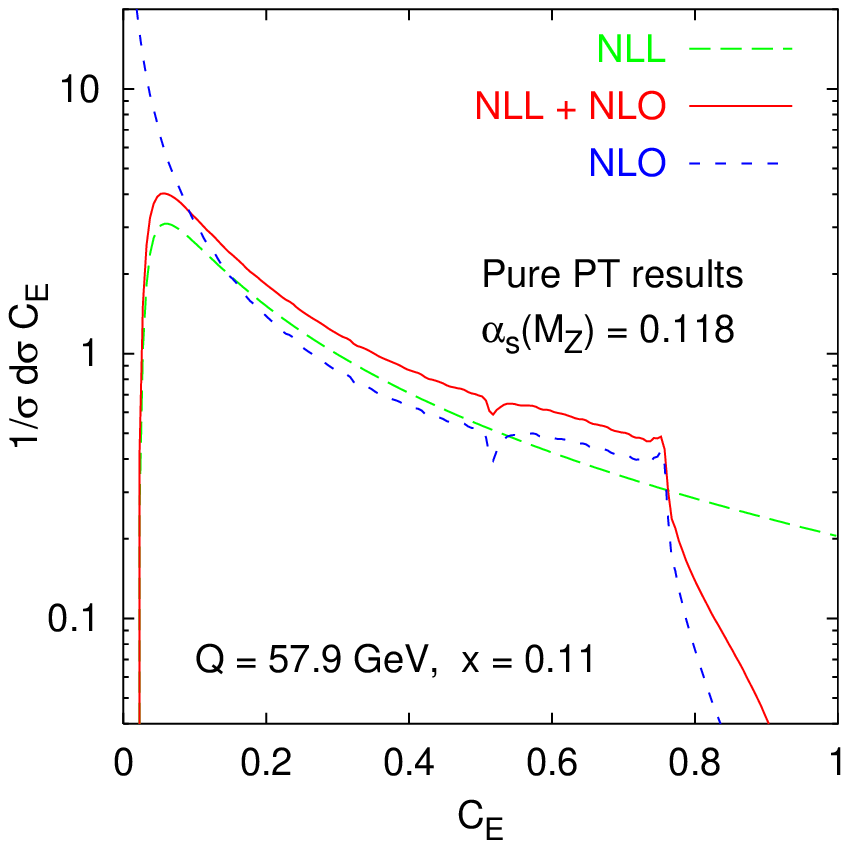,height=0.48\textwidth}
\caption{A comparison of resummed results (with and without matching
  to NLO) to pure NLO results, for the jet mass and $C$-parameter. No
  non-perturbative corrections are included.\label{fig:ResumEffect}}}

In figure~\ref{fig:ResumEffect} we show pure NLO predictions, pure NLL
predictions (with modified logs) and matched NLO-NLL results ($\ln R$
scheme, as described in \cite{dassalbroad}) for the distributions of
two observables: $\rho_E$ and the $C_E$-parameter.  Aside from leading
to the appearance of a Sudakov peak in the exclusive limit, the
resummation has a non-negligible ($20\%$-$40\%$) effect even for
intermediate values of the observable, where the logarithms being
resummed, while not formally large, can still be numerically
significant. In many cases this will be the region used for fits.

It is interesting also to examine the region where the value of the
observable is large. For the jet mass we see that the matched-resummed
curve coincides almost exactly with the NLO curve close to the
kinematical limit $\rho_E \lesssim 1/4$, where the leading
contribution is $\order{\as}$. The situation for $C_E$ is more
complicated because beyond $C_E = 3/4$ the leading contribution is
$\order{\as^2}$. There one sees that the NLL$+$NLO curve behaves very
differently from the pure NLO curve. This is almost certainly an
artefact of the matching procedure, which is not intended to be used
in situations where the leading term is not of order $\as$.

\subsection{Choice of default $X$-scale for the logs}

In a resummation which is truncated to some fixed logarithmic order
(\eg NLL) there is an ambiguity associated with the choice of
logarithm to be resummed. For example for the jet mass we resummed
$\ln 1/\rho_E$, while for the thrust we took $\ln 2/\tau_{tE}$. The
different factors were motivated by the kinematical relation between
the jet mass and the thrust, derived in section~\ref{sec:kinematics}.
But we could just as easily have chosen to resum $\ln 2/\rho_E$ and
$\ln 1/\tau_{zE}$ --- the single-logarithmic contribution in terms of
these redefined logarithms would have a different functional form, but
the overall answer would be the same to NLL accuracy.

In \cite{dassalbroad} we formalised this ambiguity by introducing an
``$X$-scale'' parameter, allowing us to redefine the logarithm to be
resummed,
\begin{equation}
  L = \ln \frac{V_0}{V} \longrightarrow {\overline L} = \ln \frac{V_0}{XV}\,,
\end{equation}
where $V_0$ is the default coefficient in the numerator of the
logarithm ($\rho_{E,0}=1$, $\tau_{tE,0} = 2$, $C_{E,0} = 12$).

Having introduced the parameter $X$ one should establish a consistent
procedure for setting it for a range of variables. We adopt the
convention of choosing the default $X=X_0$ such that the resulting
${\overline G}_{11}$ coefficient only contains terms associated with hard
collinear contributions (the yellow bands in fig.~\ref{fig:discont}).
For $\tau_{tE}$, $\tau_{zQ}$, $\rho_E$ and $C_{E}$ (as for all the
usual $\ee$ resummations) this corresponds to $X_0=1$, while for
$\tau_{zE}$, $X_0=1/2$ and for $B_{zE}$, $X_0 = 1/\sqrt{2}$.

\begin{figure}[tp]
    \begin{center}
  \epsfig{file=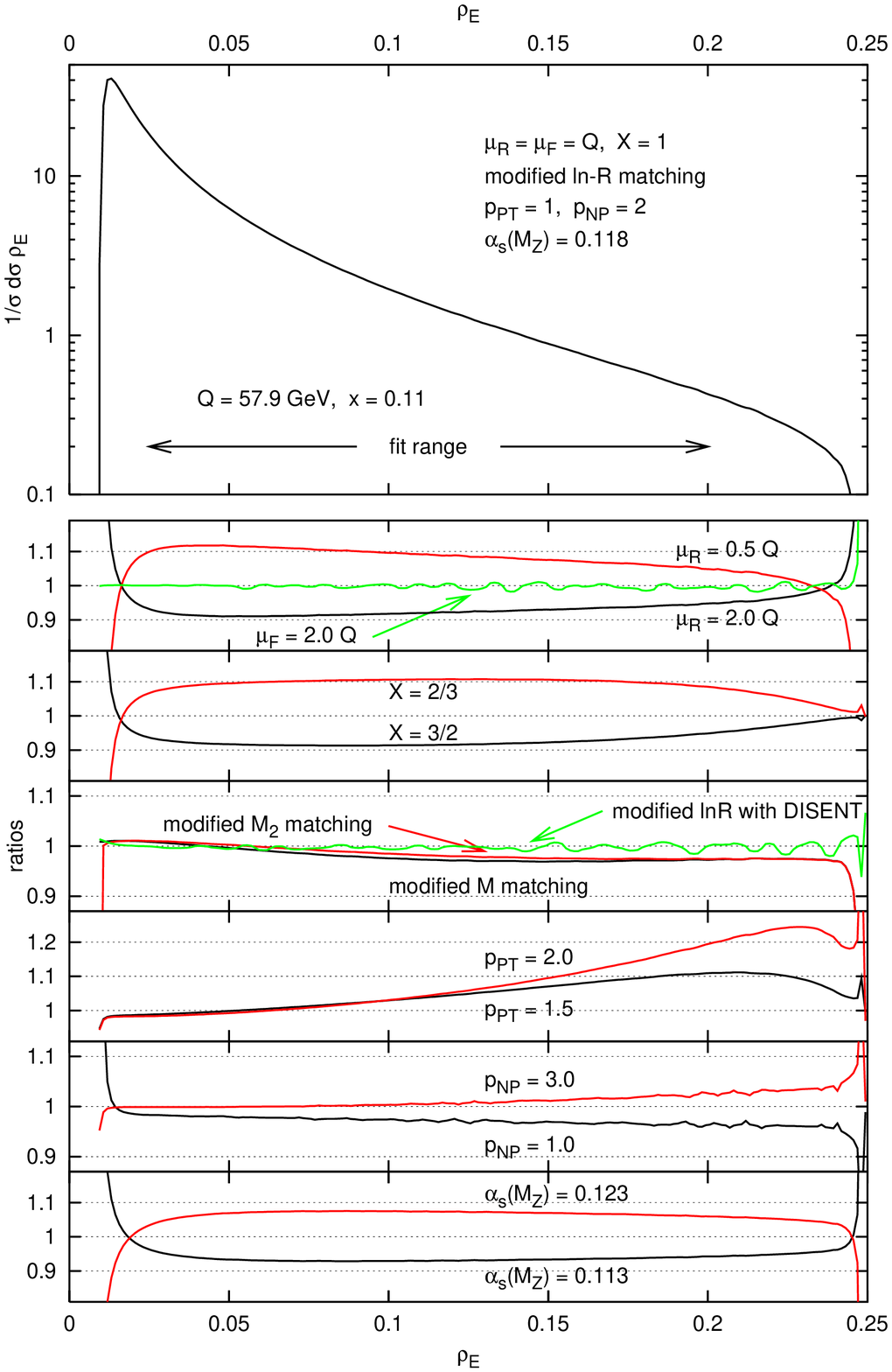,width=0.85\textwidth}
    \caption{The upper plot shows the $\rho_E$ distribution for a
      standard set of scale and matching parameters. The remaining
      plots shows ratios to this standard distribution that are
      obtained when varying different scale and matching parameters.
      The fit range indicated in the upper plot is that used later, in
      section~\ref{sec:data}. }
    \label{fig:rhoUncert}
    \end{center}
\end{figure}

\begin{figure}[tp]
    \begin{center}
  \epsfig{file=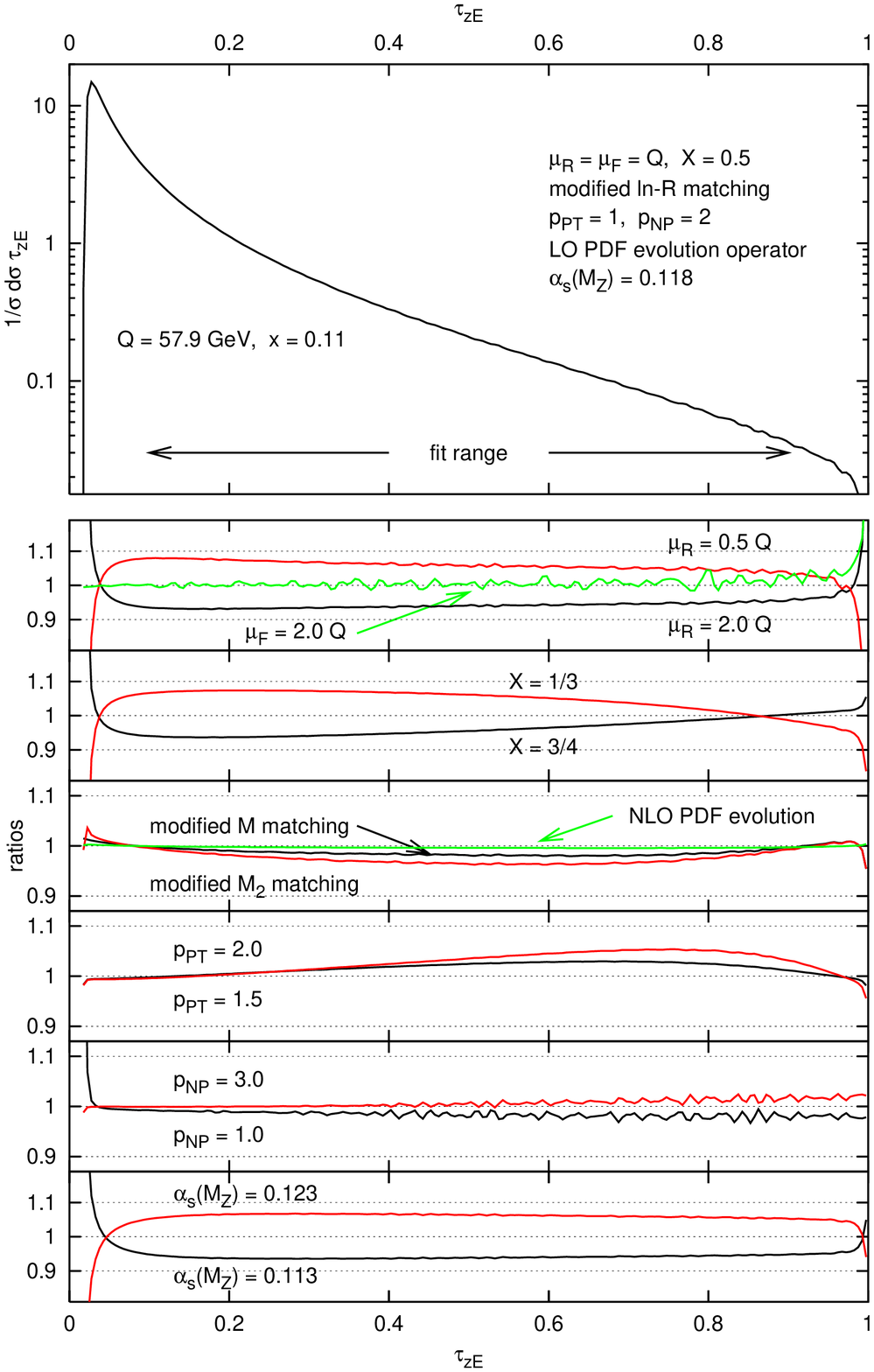,width=0.85\textwidth}
    \caption{As in figure~\ref{fig:rhoUncert}, but shown instead for
      $\tau_{zE}$. Note that the vertical scales are different in some
      of the plots.}
    \label{fig:tauzEUncert}
    \end{center}
\end{figure}



\subsection{Summary of uncertainties}
\label{sec:uncertainties}

There are a number of potential sources of uncertainty on matched
resummed event-shape distributions, essentially associated with
unknown higher-order contributions.

We can obtain an indication of the size of higher-order corrections in
a number of ways, and the effects are illustrated in
figs.~\ref{fig:rhoUncert} and \ref{fig:tauzEUncert}. These figures
show the matched resummed distributions ($\rho_E$ and $\tau_{zE}$
respectively) for default choices of parameters, matching procedures
and non-perturbative corrections, together with the ratio to the
default distributions when varying different scale and
matching parameters.

The first ratio plot shows the effect of varying $\mu_{R}$ by a factor
of two, with the effect being at about the $10\%$ level, and slightly
larger for $\rho_E$. Also shown is the effect of taking $\mu_F=2Q$
(with $\mu_R = Q$), which is in comparison negligible --- it should be
kept in mind of course that at this $x$ value the scale dependence of
the structure function is quite weak, however it is generally true
that the resummation is much more sensitive to $\mu_R$ than to
$\mu_F$.

The next plot shows the effect of varying $X$. The range to choose for
$X$ is somewhat arbitrary (as is the range for the renormalisation
scale). The question has been investigated in some detail in
\cite{LEPQCD} for $\ee$ event shapes, where the effect of variations
in $X$ has been compared analytically to known higher-order effects,
for example to the size of the subleading $\asb^2 L$ ($G_{21}$) term.
As would be expected a suitable range is somewhere between $1/\sqrt2$
to $\sqrt2$ and $1/2$ to $2$, with the actual range in \cite{LEPQCD}
chosen to be between $2/3$ and $3/2$. This particular range has the
feature in $\ee$ event shapes that the overall impact of the $X$
variation is similar to that of the (standard) $\mu_R$ variation, the
purpose of the $X$ variation then in part being to have access to an
alternative functional form for the uncertainty.  This is also the
range used in figs.~\ref{fig:rhoUncert} and \ref{fig:tauzEUncert} and
we note that here too, the effect is of the same order as that from
$\mu_R$ variation, though the functional dependence on the observable
is somewhat different.

Next we investigate the effect of different kinds of matching,
described in \cite{dassalbroad}, related to analogous procedures in
$\ee$ \cite{CTTW}. In the default $\ln R$ matching ($R$ is given by
eq.~\eqref{eq:result}) $\ln R$ is
adjusted to have the correct $\order{\as^2}$ expansion and then
exponentiated to give the integrated distribution. In $M$ matching,
one calculates the difference between the expansion of $R$ and
the exact distribution at first and second order in $\as$, and this
difference (multiplied by $\Sigma_q$, so as to ensure correct behaviour
in the $V\to0$ limit, essentially the condition that $dR/dV \to
0$ for $V\to 0$) is then added to $R$. Actually the difference
between first order resummed and exact distributions will
automatically go to zero for $V\to0$ --- it is only at second order
that this does not happen (because we do not know the values of the
$G_{21}$ and 
$C_{2}$ terms of the resummation: in $\ee$ these are usually fitted
to the fixed order, but this is not feasible in DIS). Accordingly the
multiplication by $\Sigma_q$ is only really necessary at second order,
which then gives $M_2$ matching. The plot shows the effect of changing
to $M$ and $M_2$ matchings, and we see that it is small.

Though not strictly an uncertainty, for $\rho_{E}$ we show
additionally the effect of replacing the fixed order calculation from
\disaster with that from \disent. The effect turns out to be
negligible. The same is found to be true for most of the other
observables as well, despite the fact that an examination of the
logarithmic structure does in general reveal differences at
$\order{\as^2 \ln^3 V}$. There are two main reasons why the the effect
of changing to \disent is relatively small. Firstly the largest
differences between \disent and \disaster are in the gluon induced
channel, which is relatively suppressed at this $Q$ value. Secondly
because of the structure of $\ln R$, $M$ and $M_2$ matching, any
differences between the 
\disent distribution and the correct logarithmic
structure are multiplied by a Sudakov form factor, and so are
suppressed by the matching procedure.\footnote{We note that in
  $R$-type matching procedures (not easily used given the currently
  available tools in DIS), one finds significant differences at lower
  $Q$ and lower values of the observables.} Accordingly in some
instances in this paper, where for technical reasons we did not have
suitable \disaster fixed-order results, we use instead \disent results
(one such case is in the estimates of the $\mu_F$ dependence).

In the case of $\tau_{zE}$, which involves the evaluation of the PDFs
at a scale $\sqrt{\tau_{zE}}Q$ we show the effect of changing
from LO to NLO PDF evolution (as explained in \cite{dassalbroad}, at
single-log accuracy both are equally valid, though for the matching
and $C_1$ terms to be correct, the starting distribution at scale
$\mu_F$ \emph{must} be an NLO parton distribution). At this value of
$Q$, $\langle x \rangle=0.11$ and scaling violations are small, and as
a result there is very little effect in switching to NLO PDF
evolution. At other $Q$ values (with correspondingly different $x$
values), the effects are visible, but still relatively unimportant.

Another aspect of matching is the use of modified logs,
\begin{equation}
  {\widetilde {\overline L}} = \frac{1}{p_{PT}} \ln \left(\frac{V_0^{p_{PT}}}{XV^{p_{PT}}} -
    \frac{V_0^{p_{PT}}}{XV_{\mathrm{max}}^{p_{PT}}} + 1\right)\,,
\end{equation}
where, in the original formulation \cite{CTTW} $p_{PT}$ was 1. This
modification of the logarithms is necessary to ensure that $R(V)$ 
goes to the correct value (total cross section minus cross
section for $\ec < \elim$) at $V_\mathrm{max}$ and in
some cases additionally that the distribution goes to zero at
$V_\mathrm{max}$. This modification introduces terms involving
products of $V^{p_{PT}}$ and $\ln V$. Such terms are known to exist
(see \eg \cite{CTW}).  However only terms $V^n\ln^m V$ with $n
\ge 1$ are expected to be present, so we restrict ${p_{PT}} \ge 1$. Two
possible values $p_{PT} > 1$ are shown in figs.~\ref{fig:rhoUncert}
and \ref{fig:tauzEUncert}, and in the region of small variable, the
$p_{PT}$ dependence is fairly weak, though it increases, especially
for $\rho_E$ towards the upper kinematical limit.

Analogously, one can modify the power correction so that it vanishes
at the upper limit of the observable. This is especially relevant for
observables whose distribution remains finite at the upper limit. The
modification that we choose \cite{dassalbroad} is 
\begin{equation}
  \delta V_{NP} \to  \left(1 -
    \left(\frac{V}{V_{\mathrm{max}}}\right)^{p_{NP}}\right)\delta V_{NP}\,.
\end{equation}
By default we have taken $p_{NP}=2$ (on the grounds that we wish to
modify the simple picture of a shift as little as possible). We see
that the effect of varying it between $1$ and $3$ is small.
\clearpage

For reference, the last ratio plot shows the effect of varying $\as$
by $\pm 0.005$. It enables one to make an order of magnitude estimate of
the relation between uncertainties on the distributions and a
corresponding uncertainty in a fit for $\as$. We point out however
that when carrying out a simultaneous fit for both $\as$ and
$\alpha_0$, the situation is more complicated because of the
correlation between the two quantities.


\section{Comparison to data}
\label{sec:data}

\TABLE{
    \begin{tabular}{|r|r|r|l|}
      \hline
      $Q_\mathrm{min}$ & $Q_\mathrm{max}$ &
      $\langle Q \rangle$ & $\langle x \rangle$
      \\ \hline
      $ 7$ & $  8$ & $ 7.5$ & $0.0039$ \\
      $ 8$ & $ 10$ & $ 8.7$ & $0.0060$ \\
      $14$ & $ 16$ & $15.0$ & $0.014 $ \\
      $16$ & $ 20$ & $17.8$ & $0.020 $ \\
      $20$ & $ 30$ & $23.6$ & $0.031 $ \\
      $30$ & $ 50$ & $36.7$ & $0.056 $ \\
      $50$ & $ 70$ & $57.9$ & $0.11  $ \\
      $70$ & $100$ & $81.3$ & $0.18  $ \\\hline
    \end{tabular}
    \caption{The bins for the H1 data \cite{H1NewData}. $Q$ values are
      given in GeV.}
    \label{tab:Qxvals}
}
In this section we fit a range of resummed DIS event-shape
distributions to H1 data \cite{H1NewData}. It would be beyond the
scope of our expertise to carry out a `proper' fit, taking into
account for example full correlations of the systematic errors. Rather
the results of this section should be taken as indicative, and in
particular the errors on our fit results are likely to be
underestimated.\footnote{For those wishing to carry out a more
  sophisticated analysis, the programs used to calculate the matched
  resummed distributions are available from
  \texttt{http://www.lpthe.jussieu.fr/\~~\!\!\!salam/disresum/}.}

We shall examine all the measured event shapes, with the jet mass
taken in the $p$ scheme, as discussed in
section~\ref{sec:masseffects}.\footnote{We thank Uli Martyn for
  providing us with the numerical values for the distribution of this
  observable.} The data is binned in $Q$ as shown in
table~\ref{tab:Qxvals}.

Formally when calculating the theoretical distribution, one ought to
divide the $Q$ and $x$ ranges into small sub-bins, calculate the full
matched and non-perturbatively shifted distribution in each sub-bin
and then recombine to obtain results for the whole bin. The main
practical limitation on this approach is that of calculating the
fixed-order perturbative distributions for all the sub-bins, which
even with the methods of section~\ref{sec:speedyMC} would probably
have required computing resources beyond those available to
us.\footnote{The problem could probably be rendered tractable using an
  adaptation of methods described in \cite{Kosower}.} Instead we
simply calculate the distributions for the mean $x$ and $Q$ values as
given in table~\ref{tab:Qxvals}. Both the theoretical and experimental
distributions are normalised to the total cross section for the energy
in the current hemisphere to be larger than $\elim = 0.1 Q$.

\FIGURE{\epsfig{file=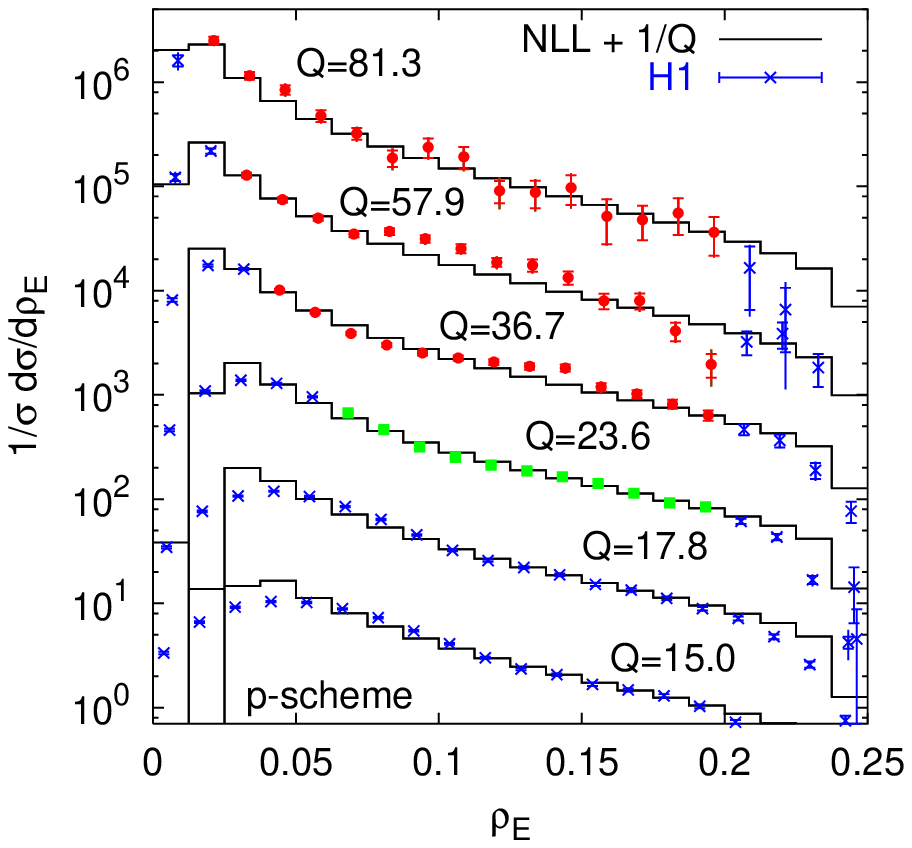,width=0.55\textwidth}\vspace{-0.7cm}
  \label{fig:ladderrho}
  \caption{H1 data for $\rho_E$ compared to the fitted theoretical
    predictions. For more details, see the caption of
    figure~\ref{fig:4ladders}.}
  }

The fits will involve two free parameters: $\as(M_Z)$ and the
non-perturbative parameter $\alpha_0(\mu_I)$ with $\mu_I=
2\,\mathrm{GeV})$. We aim to learn several things from them: we wish
to verify the consistency of $\alpha_0$ (and $\as$) with other
measurements (from $\ee$ means and distributions and DIS means). And
we would also like to check that the fit results are consistent across
a range of $Q$ values.  To accomplish this we will carry out two sets
of fits, one for all data with $Q > 30\,\mathrm{GeV}$ and a second one
with additionally the bin $20 > Q > 30\,\mathrm{GeV}$. The $20 > Q >
30\,\mathrm{GeV}$ region on its own has a statistical weight slightly
larger than the combined $Q > 30\,\mathrm{GeV}$ region.

The upper limit on the variable in the fit is chosen to be below the
$\order{\as}$ maximum value for the observable (which in some cases
coincides with the all-orders maximum), because beyond this point the
fixed-order distribution is known only at leading order and because
the current matching procedures make little sense in steeply falling
regions beyond the $\order{\as}$ maximum. The lower limit on the
variable (applied to the bin centre) is chosen to scale as $1/Q$. The
coefficient of the scaling is taken larger for observables with larger
power corrections. In some cases at the highest $Q$ point even the
lowest variable-bin is included. While at first sight it may look like
one is therefore probing the distribution down to zero (which would be
unsafe), what one is actually doing in such a bin is probing the
integrated distribution at the upper boundary of the bin, which is
potentially acceptable as long as that upper boundary is in the immediate
neighbourhood, or to the right of the maximum of the distribution.

The fitted distributions are compared to data in
figures~\ref{fig:ladderrho} ($\rho_E$) and \ref{fig:4ladders} ($C_E$,
$\tau_{tE}$, $\tau_{zE}$ and $B_{zE}$).\footnote{We note that for the
  $C$-parameter we have used \disent rather than \disaster fixed-order
  predictions --- this is for technical reasons and in light of the
  discussion in section~\ref{sec:uncertainties} it is not expected to
  affect the results significantly.} Only the red (round) points have
been used in the fit (the green squares correspond to those points
that will additionally be used in the fit for $Q > 20$~GeV). The
$1$-sigma contours of the fit results are shown in
figure~\ref{fig:contours}a and the numerical results together with the
$\chi^2$ values are given in table~\ref{tab:chi2}.

One sees remarkably good consistency between the observables, with the
possible exception of the broadening, whose $\as$ value exceeds the
world average by just over a standard deviation (but one should keep
in mind that our neglect of correlations between errors may lead to an
underestimate of the errors). The results are also in very good
agreement with those from $\ee$ mean values (see for example the
$p$-scheme results of \cite{SalamWicke}). Furthermore they are reasonably
compatible with results from fits to means in DIS \cite{H1NewData,ZEUS} and
from distributions in $\ee$ \cite{UniversalityTest}, though in these two
cases the internal consistency between fits to different observables
is not perfect.

\FIGURE{\mbox{}\hspace{-0.02\textwidth}
  \epsfig{file=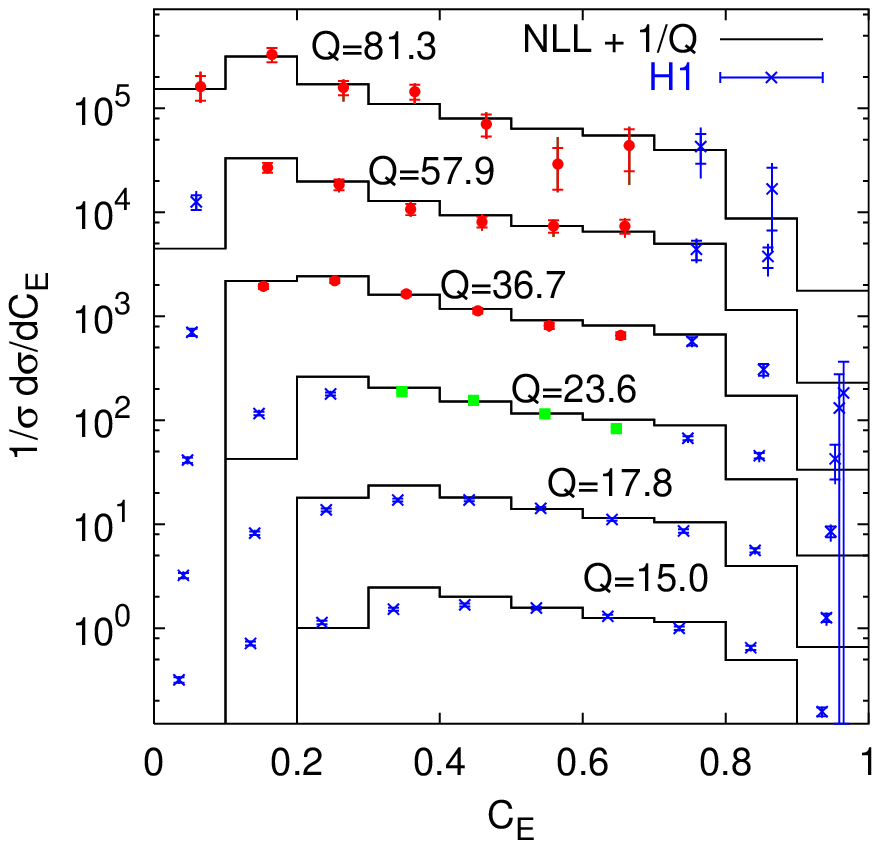,height=0.47\textwidth}
  \hspace{0.00\textwidth}
  \epsfig{file=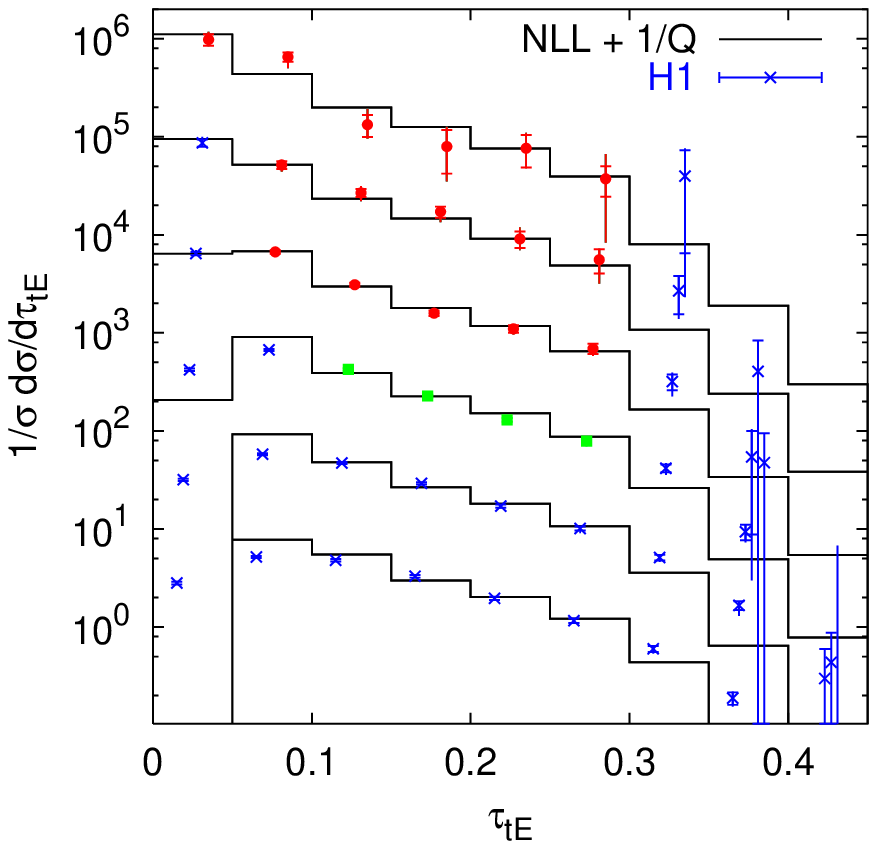,height=0.47\textwidth}
  \vspace{0.3cm}\\\mbox{ }\hspace{-0.027\textwidth}
  \epsfig{file=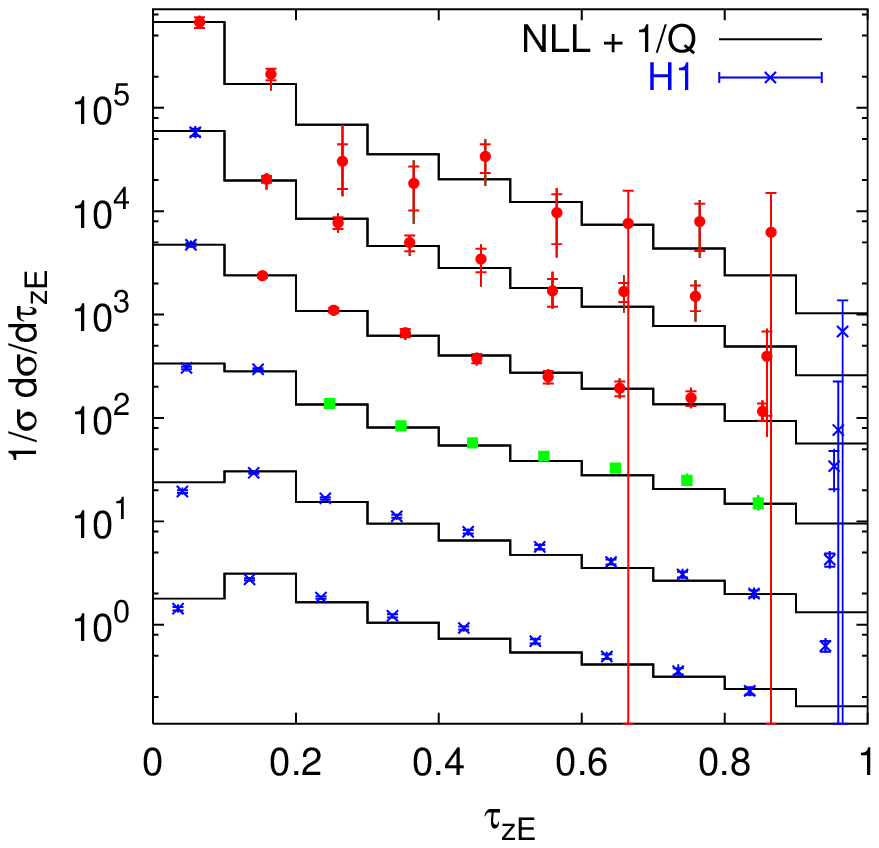,height=0.47\textwidth}%
  \hspace{0.020\textwidth}%
  \epsfig{file=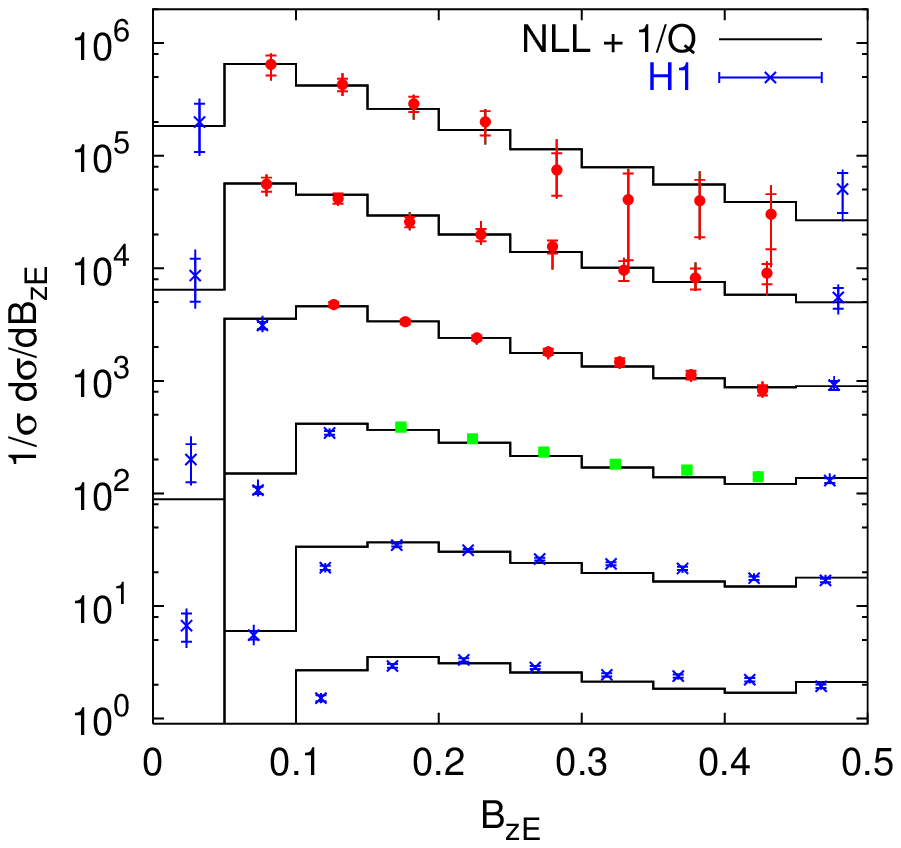,height=0.47\textwidth}
  \label{fig:4ladders}
  \caption{H1 data for $C_E$, $\tau_{tE}$, $\tau_{zE}$ and $B_{zE}$,
    compared to theoretical distributions, fitted separately for each
    observable using data with $Q > 30$~GeV (red (darker-grey in B\&W)
    round points only). In the fits for $Q > 20\,\mathrm{GeV}$,
    discussed in the text, the green (lighter-grey in B\&W) square
    points are also used.  For clarity, points in the same
    variable-bin but at different $Q$ values are staggered.}  }

\FIGURE{\epsfig{file=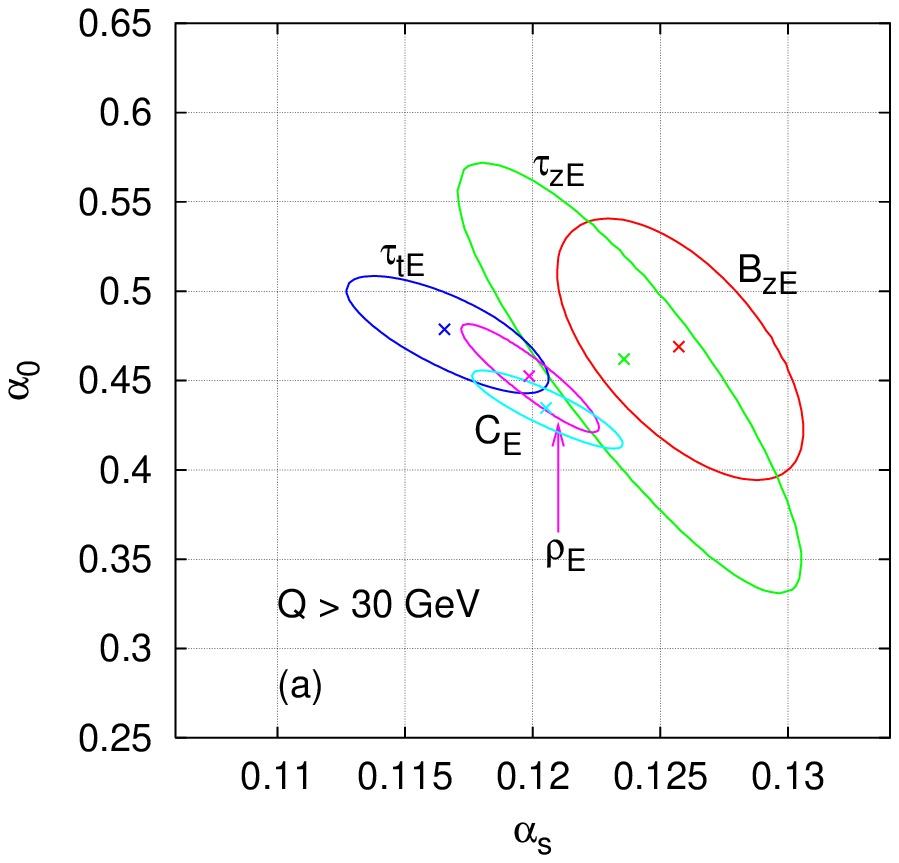,width=0.47\textwidth}
  \hfill
  \epsfig{file=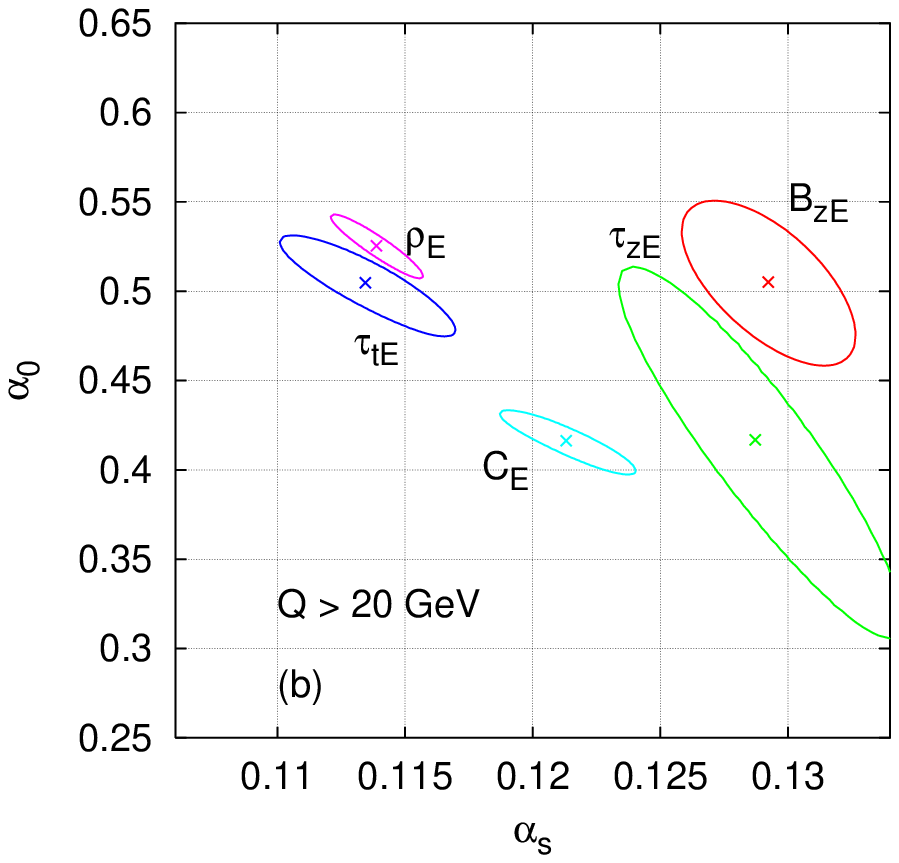,width=0.47\textwidth}
  \label{fig:contours}
  \caption{$1$-$\sigma$ contours for fits to the five event shapes
    measured by H1. Asymmetric systematic errors have been averaged
    and added in quadrature to the statistical errors. For plot (a),
    the points fitted are those shown as red circles in
    figures~\ref{fig:ladderrho} and \ref{fig:4ladders}, while the fit
    for plot (b) includes additionally the green points (squares) of
    those figures.}}

Returning to the DIS distributions, for all variables except $\rho_E$
the $\chi^2$ results are reasonable.  Looking at the comparison
between the $\rho_E$ data and theoretical predictions,
figure~\ref{fig:ladderrho}, there seem to be problems in the central
region of the distribution for for the $\langle Q\rangle = 57.9$~GeV
bin and also somewhat for the $\langle Q\rangle = 36.7$~GeV bin. There
is no evidence for any such discrepancy in the $Q$ bins immediately
above and below, so it is a problem which needs to be better
understood.

\TABLE[t]{
  \begin{tabular}{|c |c| c| r|} \hline
    $\cV$       & $\as$ & $\alpha_0$& $\chi^2/$d.o.f.  \\ \hline
    $\rho$      & $0.1199\pm0.0027$ & $0.452\pm0.030$& $90.4/40$ \\ \hline
    $C_E$       & $0.1205\pm0.0029$ & $0.435\pm0.021$& $25.7/17$ \\ \hline
    $\tau_{tE}$ & $0.1165\pm0.0039$ & $0.479\pm0.032$& $11.9/14$ \\ \hline
    $\tau_{zE}$ & $0.1236\pm0.0067$ & $0.462\pm0.121$& $11.0/23$ \\ \hline
    $B_{zE}$    & $0.1257\pm0.0048$ & $0.469\pm0.073$& $6.7/21$  \\ \hline
  \end{tabular}
  \label{tab:chi2}
  \caption{Fit results and $\chi^2$ values using data with
    $Q>30$~GeV. }}

It is interesting also to examine what happens when we extend our fits
to lower $Q$.  Figure~\ref{fig:contours}b shows the fit results when
one additionally includes the $\langle Q \rangle = 23.6$~GeV bin (the
green points of figures~\ref{fig:ladderrho} and \ref{fig:4ladders}).
Because of the larger number of events at lower $Q$ values, the
statistical weight of the $\langle Q \rangle = 23.6$~GeV data points
is comparable to and sometimes even larger than that from all the
higher-$Q$ bins combined. Its inclusion leads to a significant
reduction in the consistency between different observables, though
$\alpha_0$ is still variable-independent to within $\pm 10\%$.

The decrease in consistency as one one goes to lower $Q$, both between
observables and compared to the world average for $\as$, could well be
due to the relatively more important role of higher order
contributions at lower $Q$, whether in the form of $\as(Q)/Q$ or
$1/Q^2$ non-perturbative effects or NNL perturbative contributions.
It would be of considerable interest to have higher-precision data at
the larger $Q$ values so as to be able to trace such effects in a
continuous fashion and perhaps even identify their origin. 

In this context we note that the apparent more limited consistency
between observables in fits to DIS mean values \cite{H1NewData,ZEUS}
might be due to the inclusion of very low $Q$ data. To establish
whether this is truly the explanation, there is a need for higher
statistics in order to be able to carry out a meaningful fit using
just the higher-$Q$ mean values.

A point of some interest is that it is the observables measured with
respect to the photon axis that are less consistent with the world
average for $\as$ (giving larger values). The fact that these
observables are directly sensitive to the transverse momentum of the
incoming parton leads one to wonder whether we are seeing the early
onset of some small-$x$ effect or of an `intrinsic' transverse
momentum of the proton. An extension of measurements and theoretical
predictions to a wider range of observables might therefore be of
interest, in particular for observables which exist in `thrust-axis'
and `photon-axis' variants (in analogy to the two thrusts), such as
the broadenings (\ie $B_{tE}$ as well as $B_{zE}$) and the thrust
majors. Theoretical predictions for the latter would probably need to
use the semi-numerical methods developed in \cite{BanfiSalamZander}.

\TABLE{
\begin{tabular}{|c|r|r|c|c|}
\hline
 & \multicolumn{2}{|c|}{pure NLO} & 
\multicolumn{2}{|c|}{No NG logs} \\\hline
$\cV$ & \multicolumn{1}{|c|}{$\delta\as$} & \multicolumn{1}{|c|}{$\delta\alpha_0$} & $\delta\as$ & $\delta
\alpha_0$ \\ \hline
$\rho_E$    & $+0.019$  &  $+0.08$ & $+0.005$ & $+0.03$ \\ \hline
$C_E$       & $>+0.030$ & $<-0.14$ & $+0.008$ & $+0.01$ \\ \hline
$\tau_{tE}$ & $+0.019$  &  $+0.05$ & $+0.006$ & $+0.02$ \\ \hline
$\tau_{zE}$ & $+0.008$  &  $+0.01$ & $+0.001$ & $+0.03$ \\ \hline
$B_{zE}$    & $+0.012$  &  $-0.29$ & ---      & ---      \\ \hline
\end{tabular}
\caption{Changes in the fit results for $\as$ and $\alpha_0$ if (left)
  we fit with just pure NLO predictions (plus power corrections)
  rather than matched NLL$ + $NLO; and (right) if we fail to account
  for non-global logarithms in the resummed part of the
  result.\label{tab:IllegalFits}} }

As an aside, for illustrative purposes we also examine what impact the
resummations themselves have on the fit results. In the left-hand part
of table~\ref{tab:IllegalFits} we show the change in the $\as$,
$\alpha_0$ fit results (with $Q>30$~GeV) where the matched resummed
perturbative part has been replaced by the pure NLO result, while
keeping the fit-range fixed.  The $\as$ results clearly become
incompatible with the world average, at about the $15$--$20\%$ level,
though for all variables other than $C_E$ (where we have technical
problems with the fit) the $\chi^2$ values remain reasonable. The
$\alpha_0$ values remain more or less consistent with our expectations
(except for $B_{zE}$). 

In the right-hand part of table~\ref{tab:IllegalFits} we show the
changes in $\as$ and $\alpha_0$ that follow from neglecting non-global
logarithms in the resummed component of matched, power-corrected fits
to the data. One sees effects of between $5\%$ and $7\%$ on $\as$ for
non-global variables and much less of an effect for the
discontinuously global $\tau_{zE}$. The non perturbative $\alpha_0$
parameter is in contrast relatively insensitive to non-global effects.
It should be kept in mind that in these fits the matching in any case
fixes up the $\as^2 L^2$ term, so it is only the effect of terms
$\as^n L^n$ with $n \ge 3$ (about half the total effect) that is
reflected.

Finally, returning to the normal matched resummed fits, we should
emphasise that we have not considered the impact of the uncertainties
discussed in section~\ref{sec:uncertainties}. These should be examined
in future studies, together with an analysis of the stability of the
results with respect to variations in the fit ranges (with the current
statistics this is difficult unless one includes lower-$Q$ data), and
of the impact of integrating the predictions over the whole bin rather
than just taking them at the mean $x,Q$ of the bin. Finally if and
when the tools for fixed-order calculations including $Z$ exchange and
interference become available, then it would be useful to evaluate
their impact on the results (the resummations themselves are not
affected, but the constant terms and matching are).




\section{Conclusions}
\label{sec:conclusions}

The study of DIS event shapes has been largely motivated by the vast
amount of HERA data that has become available over the past few years
and the notable success of a similar study for $\ee$ variables.  While
mean values were the first properties of DIS event shapes to be
examined theoretically,
interest is shifting to the study of differential distributions for
which there are over 500 data points available from H1. Here we shall
summarise the developments that have taken place during the entire
course of our study involving DIS event shape distributions.

The project of studying the DIS distributions implies the calculation
of resummed predictions, matched to fixed order ones to enable one to
study the distribution over a wide enough range of values. While it
was initially believed that the DIS resummed predictions would offer
relatively few extra complications, given the techniques in place for
$\ee$ variables (the development of which spanned nearly a decade) and
the similarities in definitions of the event shapes, this turned out
not to be the case.

From a purely theoretical viewpoint the first complication encountered
was that of the different treatment, of collinear logarithms
associated with the parton distribution functions and the presence of
an incoming gluon channel which features were naturally absent in the
$\ee$ cases. In \cite{ADS,dassalbroad} we demonstrated that for boson
axis variables such collinear radiation (including the incoming gluon
channel) is associated with a change in scale of the structure
function $q(x,Q^2) \to q(x,V^n Q^2)$ where $V$ denotes the variable.
This variable dependence of the scale of the structure function 
had been encountered prior to work on DIS event shapes in for example 
the study of the transverse momentum distribution 
of a vector boson produced in hadron-hadron collisions (see e.g \cite{DYpair}).More recently it was also established to be present 
in certain three jet event shapes, both 
in DIS and hadron-hadron collisions, for which 
we refer the reader to the 
last two items of 
\cite{MilanMultiJet}. 
In contrast for two (1+1) jet DIS 
event-shape variables not defined using the 
boson axis, such as those
discussed in the present article, one just needs to factorise the
parton densities as for a cross section and the relevant scale of the
PDFs remains $Q^2$.

While straightforward to understand theoretically, the presence of
parton distributions, different incoming partonic channels, transverse
and longitudinal components of the result with their different
$y_{Bj}$ dependences and the presence of anomalous dimension matrices
rather than simple numbers in the coefficients of certain single
logarithmic terms, creates considerable problems of notation and
particularly implementation. We dealt with and explained most of these
issues in Refs.~\cite{ADS,dassalbroad} and therefore in this article
presented the results in only a skeletal form.

One other problem that we encountered during the course of our study
was the lack of a suitable fixed order Monte Carlo program with which
to combine (match) our results.  While both \disent and \disaster were
available to us, we found that our analytical resummations, expanded
to NLO, indicated disagreement, in certain terms, with \disent and
compatibility with \disaster. We would therefore have preferred to use
\disaster for our purposes but for the fact that it was slow. We
therefore developed a package (dispatch), described in Sec.~6, that
considerably speeds up (by an order of magnitude) \disaster, making it
possible to obtain results for several $x,Q$ points with a precision
which, from the point of view of computing time, would have been
unfeasible previously.

While on the subject of fixed order computations we also developed
several matching schemes to combine our results with those from NLO
Monte Carlos \cite{dassalbroad}. While some of these (e.g $\ln R$)
involved extensions and generalisations of the corresponding $\ee$
procedures \cite{CTTW} (albeit more subtle in the DIS case) we also
proposed, for the first time, multiplicative matching schemes ($M$ and
$M_2$ matching) \cite{dassalbroad}.

Another development that we made in the course our study of DIS event
shapes was the writing of our own PDF evolution code. This was
necessitated by for example the need for flexibility in the evolution,
specifically to fit $\as$ using the same value for all components
of the result rather than have part of the result `contaminated' by the
$\as$ used in the standard (\eg \cite{MRST,CTEQ}) PDF sets. Also it was
found that for our particular variables, there were problems arising
from the non-smoothness of the interpolation procedure used in
\cite{MRST}.  Our PDF evolution code enabled the discovery of bugs in
the standard evolution codes (MRST and CTEQ), which have as a result
been fixed \cite{MRST01,CTEQ6}.

Technical spin-offs apart, one important theoretical discovery made
during our investigation of DIS event shapes turned out to be that of
non-global logarithms. These have been discussed in great detail
elsewhere \cite{dassalNG1,dassalNG2}, while in the current article we
have considered their extension to discontinuously global variables,
correcting (phenomenologically relatively unimportant) omissions in
\cite{ADS,dassalbroad}.

Lastly, we come to the original aim of this work: the comparisons with
data. These give a strong boost to the idea of a universal
non-perturbative extension to the standard coupling $\as$. In the
case of most variables we find that the agreement with data at
intermediate to larger $Q$ values is remarkable down to very small
values of the variables in question.  The $\as$ and $\alpha_0$
values obtained by fitting our results to H1 data are in general
consistent with the world average and $\ee$ values respectively.  From
the viewpoint of theory, the noticeable worsening of the description
that occurs at low $Q$ values is an area that may require further
investigation. On the experimental front, data from ZEUS is eagerly
awaited as is high luminosity H1 data. This apart, a proper error
analysis in the fits to existing data should also be carried out.
Lastly the impact of the various subleading uncertainties involved in
our resummation on actual fits to the data should also be examined.

\acknowledgments

We wish to thank Vito Antonelli for discussions during the initial
stages of this paper. One of us (GPS) also wishes to thank members of the
LEP QCD working group for discussions
related to $X$-scale dependence. The fixed-order calculations used
here were obtained in part using computer facilities at CERN and at
the universities of Milano and Milano-Bicocca.

We also wish to thank Uli Martyn, Klaus Rabbertz and Thomas Kluge for
helpful comments and for providing us with the H1 data in numerical
form, and Yuri Dokshitzer, Einan Gardi and Johann Rathsman for
numerous stimulating discussions.

\appendix
\section{Resummed results}

The results for the quark and gluon jet mass cross-sections 
$\Sigma_q$ and $\Sigma_g$ were first derived in \cite{CTTW}. 
They can be expressed as below 
\begin{equation}
\Sigma_q(\as,L) = \int_0^{e^{-L}} J_{q}\! \left(\as,\frac{k^2}{Q^2}\right)
\frac{dk^2}{k^2} =
\frac{e^{-\mathcal{R}}}{\Gamma 
\left [1+\mathcal{R}^{'} \right ]}\,,
\end{equation} 
where the ``radiator'' $\mathcal{R}$ is given to NLL accuracy by 
\begin{equation}
\mathcal{R} = -Lf_1(\lambda)- f_2(\lambda) \,,
\end{equation}
with $\mathcal{R}'$ being the derivative 
\begin{equation}
\mathcal{R}' = -\frac{\partial}{\partial L}  [{L f_1}(\as,L)] \,,
\end{equation}
where we have used $\lambda = \beta_0 \as L, \; L = \ln \frac{1}{\rho} $ and 
in the derivative $\cR'$ have dropped subleading (NNLL) 
pieces involving the derivative 
of $f_2$.

The functions $f_1$ and $f_2$ are listed below 
\begin{equation}
\label{eq:quark}
f_1(\lambda) = - \frac{C_F}{2 \pi \beta_0 \lambda} \left [ \left(1-2 \lambda \right ) 
\ln \left(1-2\lambda \right)-2 \left ( 1-\lambda \right ) \ln \left
  (1-\lambda \right ) \right ],
\end{equation}
and
\begin{multline} 
f_2(\lambda) = - \frac{C_F K}{4 \pi^2 \beta_0^2} \left [2 \ln \left 
(1-\lambda \right ) - \ln \left (1-2 \lambda \right )\right ]\\ - 
\frac{3 C_F}{4 \pi \beta_0} \ln \left ( 1-\lambda \right ) 
-\frac{C_F \gamma_E}{\pi \beta_0} \left [\ln \left (1-\lambda \right ) - \ln 
\left(1-2\lambda \right) \right ] 
\\  -\frac{C_F \beta_1}{2 \pi \beta_0^3} \left [ \ln \left (1-2\lambda \right )-2 \ln 
\left (1-\lambda \right ) + \frac{1}{2} \ln^2 \left (1- 2 \lambda \right ) 
- \ln^2 \left (1-\lambda \right ) \right ].
\end{multline}
In the above results the $\beta$ function coefficients 
$\beta_0$ and $\beta_1$ are defined as
\begin{equation}
\beta_0 = \frac{11 C_A - 2 n_f }{12 \pi}, \; \beta_1 = \frac{17 C_A^2 - 5 C_A n_f -3 C_F n_f}{24 \pi^2}\,,
\end{equation}
and the constant $K$ is given by 
\begin{equation}
K = C_A \left (\frac{67}{18}- \frac{\pi^2}{6} \right ) - \frac{5}{9} n_f\,.
\end{equation}

For the gluonic contribution $\Sigma_g$ one only requires the leading log 
piece of the answer, since this contribution is already suppressed by a factor of $\as$ which comes from the probability of having a 
hard gluon in the current hemisphere:
\begin{equation}
\Sigma_g(\as,L) = \int_0^{e^{-L}} J_{g}\! \left(\as,\frac{k^2}{Q^2}\right)
\frac{dk^2}{k^2}
 = e^{L f_{1,g}} \,,
\end{equation}
where $f_{1,g}$ is identical to the corresponding quark result $f_1$ 
\eqref{eq:quark}, provided one replaces the quark colour charge $C_F$ 
by $C_A$.  

Finally we come to the non-global resummation factor $\mathcal{S}$. 
This was parameterised as below in Ref.~\cite{dassalNG1}
 \begin{equation}
  \label{eq:MainResult}
  \mathcal{S}(\as L) \simeq \exp\left( - C_F C_A \frac{\pi^2}{3}
     \left( \frac{1 + (at)^2}{1 + (bt)^c}\right)t^2\right),
\end{equation}
with
\begin{equation}
  \label{eq:tdef}
  t (\as L) = \frac{1}{2\pi}\int^1_{e^{-L}} \frac{dx}{x} \as(xQ) = 
  \frac{1}{4\pi \beta_0} \ln \frac{1}{1 - 2\beta_0 \as L}\,,
\end{equation}
with $\beta_0$ as above and
\begin{equation}
  a = 0.85C_A\,,\qquad b = 0.86 C_A\,, \qquad c = 1.33\,.
\end{equation} 
The above parameterisation takes into account the exact leading order
$\cO{\as^2}$ result computed in \cite{dassalNG1} and all subsequent
orders in the large $\NC$ limit (\ie neglecting contributions
$\cO{1/\NC^2}$).  The above parameterisation ought to be be accurate
to within a few percent for $t < 0.7$ which corresponds to $1-2
\beta_0 \as L \geq 0.005$ (and to within a fraction of a percent
up to $t=0.4$ which remains beyond the largest values of $t$ probed
phenomenologically).

\section{Constant pieces}\label{app:ConstantPieces}
The various coefficient functions that we shall require can be written
in the notation $C_{\delta a}^b$ where $\delta =2,L$ shall denote the
transverse or longitudinal piece, the index $a = q,g $ shall signify
the incoming particle and the upper index $b=q,g$ shall denote the
hard particle in $\hr$ off which we perform the resummation to obtain
the form factor $\Sigma_b$.

One has (multiplied in \eqref{eq:result} by $\asb = \frac{\as}{2 \pi}$) 
\begin{multline} 
C_{2q}^{g} = C_F \Theta \left
    (\frac{1}{2}-\xi \right ) \left
    [3\xi^3+\frac{2\xi-5\xi^2}{2(1-\xi)}
    -\frac{1+\xi^2}{1-\xi} \ln(1-\xi) \right ] \\
         + C_F \Theta \left ( \xi-\frac{1}{2} \right )\left [
           3\xi^3+2+\xi-6\xi^2 +\frac{4 \xi-3-\xi^2}{2 (1-\xi)}-\left
             ( \frac{1+\xi^2}{1-\xi} \right) \ln \xi \right ].
\end{multline}
We find it convenient to introduce the function $C_{2q}$ which is the
whole constant piece 
(sum of contributions from current quark and gluon pieces) that appears in the one-loop calculation of the
probability of having jet mass {\it{above}} $\rho$.
From $C_{2q}$ and $C_{2q}^g$ we can extract $C_{2q}^q$ via the relation $C_{2q}^q
= C_{2q}-C_{2q}^g$
The function $C_{2q}$ is
\begin{multline}
C_{2q}(\xi) = C_F \frac{\frac{1}{2}+6\xi^3-5\xi^2}{(1-\xi)_+} \Theta
  (\xi-\frac{1}{2})+C_F (1+\xi^2)\left [ \frac{\ln(1-\xi)}{1-\xi} \right
  ]_+ \Theta \left ( \xi-\frac{1}{2} \right ) \\
-C_F\frac{1+\xi^2}{1-\xi}\ln \xi \; \Theta\left(\xi-\frac{1}{2} \right ).
\end{multline}
Analogously one has
\begin{equation}
C_{2g} = \Theta \left (\xi - \frac{1}{2} \right )\frac{1}{2} T_R \left ( \left(\xi^2+(1-\xi)^2 \right )(4 \xi -2 +2 \ln(1-\xi) 
-2 \ln \xi )+8\xi (1-\xi) (1-2\xi) )\right )\,,
\end{equation}
and 
\begin{align} 
C_{Lq} &= \Theta \left (\xi-\frac{1}{2} \right ) C_F (2 \xi - 4 \xi^2)\,, \\
C_{Lg} &= \Theta \left (\xi -\frac{1}{2} \right ) \frac{1}{2} T_R \left 
( 8 \xi (1-\xi)(1-2\xi) \right )\,, \\
C_{Lq}^g &= C_F  \Theta \left (\xi -\frac{1}{2} \right ) (2\xi(1-\xi)^2)+
           C_F \Theta \left (\frac{1}{2}-\xi \right )2\xi^3 \,.
\end{align}
The thrust variable $\tau_{tE}$ has identical coefficient functions 
to those listed above for the the jet mass. 
For the C parameter the only difference is in the piece 
$C_{2q}$ (which also affects $C_{2q}^q$) where one needs to add 
the term $C_F \frac{\pi^2}{6} \delta(1-\xi)$ to the $C_{2q}$ term listed above.
 
Note that in order to go from the coefficient functions above to the
Bjorken $x$ dependent constant pieces relevant to $\eqref{eq:result}$
we have to take $2,L$ pieces with the appropriate weights, convolute
the functions above with the parton density functions and normalise to
the Born value for $F_2$. Incoming quark $C_{\delta q}^b$ pieces need
to be convoluted with $f(x/\xi) = \sum_{q,\bar q} e_q^2\, q
\left(x/\xi \right)$ where the sum runs over quarks and antiquarks,
and convolutions are defined as
\begin{equation}
C_{\delta q}^b (x,Q^2) = x \int_x^1 \frac{d\xi}{\xi} \,C_{\delta q}^b
(\xi)\, f\! \left(  x/\xi , Q^2\right )
\end{equation}
Similarly incoming gluon pieces $C_{\delta g}$ need a convolution with
$\left(\sum_{q,\bar q} e_q^2\right) g \left ( x/\xi \right )$ where
$g$ is the gluon density. To then go to the coefficient functions used
in eq.~\eqref{eq:result} one uses relations such as
\begin{equation}
  C^g_1(x,Q^2) = \frac{C^g_{2q}(x,Q^2) + \frac{y^2}{1 +
      (1-y)^2}C^g_{Lq}(x,Q^2)}{x f(x,Q^2)}\,,
\end{equation}
where $y$ is Bjorken-$y$.

Finally we recall that as in \cite{ADS}, the coefficient functions are
given in the DIS scheme. To go to another scheme one should add to
$C_{2q}$ and $C_{2g}$ the corresponding $F_2$ coefficient functions in
that scheme.
%


\end{document}